\documentstyle[emulateapj,apjfonts]{article}

\begin{document}

\newcommand{\ctan}{$^{13}$C($\alpha$,n)$^{16}$O}
\newcommand{\ctanb}{$^{13}$C($\alpha$,n)$^{16}$O~}
\newcommand{\nean}{$^{22}$Ne($\alpha$,n)$^{25}$Mg}
\newcommand{\neanb}{$^{22}$Ne($\alpha$,n)$^{25}$Mg~}
\newcommand{\neagb}{$^{22}$Ne($\alpha$,$\gamma$)$^{26}$Mg~}
\newcommand{\nqagb}{$^{14}$N($\alpha$,$\gamma$)$^{18}$F($\beta^+\nu$)$^{18}$O~($\alpha$,$\gamma$)$^{22}$Ne~}
\newcommand{\oagb}{$^{18}$O($\alpha$,$\gamma$)$^{22}$Ne~}
\newcommand{\cdpgb}{$^{12}$C(p,$\gamma$)$^{13}$N($\beta^+\nu$)$^{13}$C~}
\newcommand{\cdpg}{$^{12}$C(p,$\gamma$)$^{13}$N($\beta^+\nu$)$^{13}$C}
\newcommand{\cdpgctan}{$^{12}$C(p,$\gamma$)$^{13}$N($\beta^+ \nu$)$^{13}$C($\alpha$,n)$^{16}$O~}
\newcommand{\ctpb}{$^{13}$C(p,$\gamma$)$^{14}$N~}
\newcommand{\msb}{$M_{\odot}$~}
\newcommand{\ms}{$M_{\odot}$}
\newcommand{\cd}{$^{12}$C}
\newcommand{\cdb}{$^{12}$C~}
\newcommand{\neonb}{$^{22}$Ne~}
\newcommand{\neon}{$^{22}$Ne}
\newcommand{\ct}{$^{13}$C}
\newcommand{\ctb}{$^{13}$C~}
\newcommand{\rbset}{$^{87}$Rb}
\newcommand{\s}{$s$}
\newcommand{\Z}{$Z$}
\newcommand{\M}{$M$}
\newcommand{\Nb}{$N$~}
\newcommand{\fb}{$f$~}
\newcommand{\N}{$N$}
\newcommand{\f}{$f$}
\newcommand{\p}{$p$}
\newcommand{\r}{$r$}
\newcommand{\permil}{$^o\!/\!_{oo}$}
\newcommand{\permilb}{$^o\!/\!_{oo}$~}
\renewcommand{\labelitemi}{}
\renewcommand{\labelitemii}{}

\title{Si ISOTOPIC RATIOS IN MAINSTREAM PRESOLAR SiC GRAINS 
REVISITED}

\author{Maria Lugaro\altaffilmark{1}, Ernst Zinner\altaffilmark{2}, 
Roberto Gallino\altaffilmark{3}, Sachiko Amari\altaffilmark{2}}

\affil{1. Department of Mathematics, Monash University, Clayton 
3168, Victoria, Australia} 

\affil{2. Laboratory for Space Physics and the Physics 
Department, Washington University, One Brookings Drive, St. Louis, 
MO 63130, USA}

\affil{3. Dipartimento di Fisica Generale, Universit\'a di Torino, 
Via P. Giuria 1, I-10125 Torino, Italy}

Submitted to the Astrophysical Journal

April 23, 1999

Revised July 19, 1999

\begin{abstract}

Although mainstream SiC grains, the major group of presolar 
SiC grains found in meteorites, are believed to have originated in the 
expanding envelope of asymptotic giant branch (AGB) stars during their 
late carbon-rich phases, their Si isotopic ratios show a distribution 
that cannot be explained by nucleosynthesis in this kind of stars. 
Previously, this distribution has been interpreted to be the result 
of contributions from many AGB stars of different ages whose initial 
Si isotopic ratios vary due to the Galactic chemical evolution of the 
Si isotopes. This paper presents a new interpretation based on local 
heterogeneities of the Si isotopes in the interstellar medium 
at the time the parent stars of the mainstream grains were born. 
Recently, several authors have presented inhomogeneous chemical 
evolution models of the Galactic disk in order to account for the 
well known evidence that F and G dwarfs of similar age show an 
intrinsic scatter in their elemental abundances.

First we report new calculations of the \s-process nucleosynthesis of 
the Si and Ti isotopes in four AGB models (1.5, 3, and 5 \msb with \Z = 
0.02; 3 \msb with \Z = 0.006). These calculations are based on the 
release of neutrons in the He intershell by the \ctb source during 
the interpulse periods followed by a second small burst of neutrons released
in the convective thermal pulse by the marginal activation of the $^{22}$Ne
source. In the 1.5 and 3 \msb 
models with solar metallicity the predicted shifts of the Si isotopic 
ratios in the stars' envelope are much smaller ($<$ 
30 \permilb for the  
$^{29}$Si/$^{28}$Si ratio and $<$ 40 \permilb for the $^{30}$Si/$^{28}$Si 
ratio; the two ratios are normalized to solar)
than the range observed in the mainstream grains (up to 180 \permil).
Isotopic shifts are of the same order as in the SiC grains for the 5 \msb 
and \Z = 0.006 models but the slope of the $^{29}$Si/$^{28}$Si vs.
$^{30}$Si/$^{28}$Si correlation line is much smaller than that of the 
grains. We also show that none of the models can reproduce the 
correlations between the
Ti and Si isotopic ratios measured in the mainstream grains as the result of
$s$-process nucleosynthesis only.

To explain the distribution of the grains' Si isotopic compositions we 
constructed a simple Monte Carlo model in which contributions from 
classic Type  
Ia, Type Ia sub-Chandrasekhar, 
and Type II supernova (SN) models of different 
masses were admixed in a statistical way to material with a given Si 
isotopic composition. For four different starting compositions (average 
composition of the mainstream grains corrected for AGB contributions, solar composition, 100 \permilb and 
200 \permilb deficits in $^{29}$Si and $^{30}$Si relative to solar) we show that, 
with the appropriate choice of two parameters, the distribution of 
the Si isotopic ratios in the mainstream grains can be successfully 
reproduced. The parameters to be adjusted are the total number of SN 
sources selected and the fraction of the material ejected from each SN 
that is mixed to the starting material. An 
upward adjustment of the supernova yield of $^{29}$Si relative to the other Si
isotopes by a factor 1.5 was also introduced. Using current 
SN yields and Galactic chemical
evolution models, this increase is necessary to achieve the Si isotopic 
ratios of the solar system.

If most mainstream SiC grains come from AGB stars that were born within 
a short time span, local heterogeneities must be the dominant cause of 
their Si isotopic variations. However, if AGB stars of different masses 
and therefore different ages contributed SiC to the solar system, the Si 
distribution of the mainstream grains reflect both the effect of Galactic 
chemical evolution of the Si isotopes and of isotopic heterogeneity at 
the time these stars were born.

\end{abstract}


\section{Introduction}

Primitive meteorites contain presolar dust grains, grains that formed 
in stellar outflows and supernova (SN) ejecta and survived not only a 
long history as interstellar (IS) grains in the interstellar medium 
(ISM), but also the formation of the solar system and conditions in the 
meteorites' parent bodies (Anders \& Zinner 1993; Ott 1993). The stellar 
origin of these grains is indicated by their isotopic compositions, which 
are completely different from those of material in the solar system and 
reflect the compositions of their stellar sources. Their study in the 
laboratory provides information on stellar nucleosynthesis, Galactic 
chemical evolution (hereafter GCE), physical and chemical properties 
of stellar atmospheres and ejecta, and conditions in the early solar 
system (Bernatowicz \& Zinner 1997; Zinner 1998).

The following types of presolar grains have been identified to date: 
diamond  (Lewis et al. 1987), silicon carbide (SiC, carborundum) 
(Bernatowicz et al. 1987), graphite (Amari et al. 1990), aluminum 
oxide (Al$_2$O$_3$, corundum), spinel (MgAl$_2$O$_3$) (Hutcheon et al. 
1994; Nittler et al. 1994) and silicon nitride (Si$_3$N$_4$) 
(Nittler et al. 1995). In addition, graphite and SiC contain tiny 
subgrains of Ti carbide 
(graphite contains also Zr and Mo carbide) that were identified by 
transmission electron microscopy (Bernatowicz et al. 1991, 1996; 
Bernatowicz, Amari, \& Lewis 1992). The carbonaceous phases diamond, 
SiC and graphite were discovered because they carry isotopically 
anomalous noble gases (Tang \& Anders 1988a; Lewis, Amari, \& Anders 
1990, 1994; Huss \& Lewis 1994a,b; Amari, Lewis, \& Anders 1995a); 
they can be extracted from meteorites in almost pure form by chemical 
and physical processing (Tang et al. 1988; Amari, Lewis, \& Anders 1994). 
This results in relatively large amounts (micrograms) of samples that 
can be analyzed in great detail. In contrast, presolar corundum and 
silicon nitride have been discovered by ion microprobe isotopic 
measurements of individual grains from chemically resistant residues 
and only a limited number of grains ($\sim$ 100 for corundum) have been 
measured to date.

Silicon carbide, graphite, corundum, and Si$_3$N$_4$ grains are large 
enough (up to several $\mu$m in diameter for corundum and Si$_3$N$_4$, 
up to $>$ 10 $\mu$m for SiC and graphite) to be analyzed individually 
for their elemental and isotopic compositions. The ion microprobe makes 
it possible to measure isotopic ratios of major and some minor elements 
in single grains down to $\sim$ 0.5 $\mu$m in diameter (e.g., Hoppe et al.
1994, 1995; Huss, Hutcheon, \& Wasserburg 1997; Travaglio et al. 1999). 
Ion microprobe isotopic measurements have been made for C, N, O, Mg, Si, 
K, Ca, and Ti on many (for the major elements on thousands of) grains  
(see, e.g., Zinner 1998). Single grain isotopic measurements, albeit on 
a limited number of grains, have also been made for Zr, Mo, and Sr by 
resonance ionization mass spectrometry (Nicolussi et al. 1997, 1998a,b,c; 
Davis et al. 1999; Pellin et al. 1999); and for He and Ne by laser 
extraction and noble gas mass spectrometry (Nichols et al. 1991, 1994;  
Kehm et al. 1996). Diamonds, instead, are too small ($\sim$ 2 nm) for 
single grain analysis and isotopic measurements have been made on ``bulk
samples'', collections of many grains. Bulk analyses have also been made 
in SiC and graphite of the noble gases (Lewis et al. 1994; Amari et al. 
1995a) and of trace elements such as Sr, Ba, Nd, Sm, and Dy in SiC (see, 
e.g., Anders \& Zinner 1993; Hoppe \& Ott 1997; Zinner 1998).

Based on their isotopic compositions, several stellar sources have been 
identified for presolar grains. Most corundum grains are believed to have 
originated in low-mass red giant (RG) and asymptotic giant branch (AGB) 
stars. This identification rests on the O isotopic ratios and inferred 
$^{26}$Al/$^{27}$Al ratios (Huss et al. 1994; Nittler et al. 1997; Nittler
1997; Choi et al. 1998; Nittler \& Alexander 1999a). Low-density graphite 
grains, a subtype of SiC grains termed X grains, and Si$_3$N$_4$ grains have 
isotopic signatures indicative of a SN origin (Nittler et al. 1995; 
Travaglio et al. 1999). A SN origin has also been proposed for a few 
corundum grains (Nittler et al. 1998; Choi et al. 1998). In addition, 
a few rare SiC grains with large $^{30}$Si excesses and low 
$^{12}$C/$^{13}$C and $^{14}$N/$^{15}$N ratios are possibly of a nova origin
(Gao \& Nittler 1997).

Most SiC grains, in particular the ``mainstream'' component, which 
accounts for $\sim$ 93\% of all SiC (Hoppe et al. 1994; Hoppe \& Ott 1997), 
are believed to come from carbon stars, thermally pulsing (TP) AGB stars
during late stages of their evolution. 
The best evidence for such an origin are the \s-process (slow neutron 
capture process) isotopic patterns displayed by the heavy elements Kr, 
Sr, Zr, Mo, Xe, Ba, Nd, and Sm  (Lewis et al. 1994; Hoppe \& Ott 1997; 
Gallino, Raiteri, \& Busso 1993; Gallino, Busso, \& Lugaro 1997;  
Nicolussi et al. 1997, 1998a; Pellin et al. 1999) and the presence of  
Ne-E(H), almost pure $^{22}$Ne (Lewis et al. 1990, 1994; Gallino et al. 
1990). The C and N isotopic compositions as well as $^{26}$Al/$^{27}$Al
ratios of individual mainstream SiC grains are by and large consistent 
with a carbon star origin  (Hoppe et al. 1994; Huss et al. 1997; Hoppe \& 
Ott 1997).

In contrast to C, N, Ne, Al and the heavy elements, the variations in the 
Si (and Ti) isotopic ratios measured in single mainstream grains cannot be 
explained in terms of nucleosynthesis in AGB stars (Gallino et al. 1990, 
1994; Brown \& Clayton 1992a). They have been interpreted to indicate that 
many stellar sources (Clayton et al. 1991; Alexander 1993), whose 
initial compositions vary because of GCE, contributed SiC grains to the 
solar system (Gallino et al. 1994; Timmes \& Clayton 1996; Clayton \&  
Timmes 1997a). However, a fundamental problem with this interpretation 
is the fact that the metallicity implied by the Si isotopic compositions 
of the mainstream grains is higher than that of the sun. This would mean 
that the grains are younger than the solar system. A solution to this 
puzzle has been proposed by Clayton (1997) who considered the possibility 
that the sun and the AGB stars that were the sources of the mainstream 
grains did not originate in the same Galactic region but changed their 
positions because of Galactic diffusion. Alexander \& Nittler (1999), 
on the other hand, used the Si and Ti isotopic compositions of the 
mainstream grains themselves (Figs. 3 and 4) to infer the metallicity of the 
ISM at the time of solar system formation. From this exercise they 
concluded that the sun has an atypical Si isotopic composition.

In this paper we will revisit the Si isotopic compositions of the 
mainstream SiC grains. In \S 2 we will first describe the isotopic 
properties of these grains in greater detail, discuss their AGB origin, 
and review previous attempts to understand their Si isotopic ratios. 
After presenting new calculations for the nucleosynthesis of Si (and Ti) 
in AGB stars (\S 3), we will present a new approach to the problems 
of the distribution of Si isotopic ratios (\S 4): instead of assuming an
average monotonic relationship of metallicity with time in the Galaxy, 
we investigate how 
heterogeneities in the Si isotopic ratios could result from fluctuations 
in the contributions from various types of nucleosynthetic sources to 
the low-mass stars that, in their AGB phase, produced the SiC grains. 
Preliminary accounts can be found in Lugaro et al. (1999a,b).

Galactic chemical evolution models of the Galactic disk that, 
although in different ways, 
deal with compositional inhomogeneities in the ISM have previously 
been presented by Malinie et al. (1993), Wilmes \& K\"oppen (1995), 
Copi (1997) and van den Hoek \& de Jong (1997). We 
will show that our approach is consistent with the 
results obtained by inhomogeneous GCE models, which, however, did 
not address the evolution of isotopic compositions.

\section{Meteoritic SiC, the mainstream component and the Si isotope
puzzle}

Among all presolar grains, SiC has been most widely studied because it 
is relatively abundant (6 ppm in the Murchison and in similar primitive 
meteorites) and is present in various classes of meteorites (Huss \& 
Lewis 1995). Ion microprobe isotopic analyses of single grains have 
revealed several distinct classes. This is shown in Figs. 1 and 2, which 
display the C, N, and Si isotopic ratios. For historical reasons the Si 
isotopic ratios are expressed as $\delta$-values, deviations in permil
(\permil) from the solar isotopic ratios of ($^{29}$Si/$^{28}$Si)$_{\odot}$ =
0.0506331 and ($^{30}$Si/$^{28}$Si)$_{\odot}$ = 0.0334744 (Zinner, Tang, \&
Anders 1989):

$\delta$$^{29}$Si/$^{28}$Si = [($^{29}$Si/$^{28}$Si)$_{meas}$/
($^{29}$Si/$^{28}$Si)$_{\odot} -$ 1] $\times$ 1000,

$\delta$$^{30}$Si/$^{28}$Si = [($^{30}$Si/$^{28}$Si)$_{meas}$/
($^{30}$Si/$^{28}$Si)$_{\odot} -$ 1] $\times$ 1000. 

According to their C, N and Si isotopic compositions five different 
groups of grains can be distinguished and these groups are indicated 
in the figures. Also indicated in the figures are the abundances of the 
different groups. 
By far the most common grains are the mainstream grains. It should 
be noted that the frequency distributions of grains in the plots of 
Figs. 1 and 2 do not correspond to their abundances in meteorites, 
but that rare grain types, located by automatic imaging in the ion 
probe (Nittler et al. 1995; Amari et al. 1996) are over-represented. 
An additional grain with extreme $^{29}$Si and $^{30}$Si excesses 
has been found (Amari, Zinner, \& Lewis 1999); its composition
($\delta$$^{29}$Si/$^{28}$Si = 2678 \permil, $\delta$$^{30}$Si/$^{28}$Si =
3287 \permil) lies outside the boundaries of the plot in Fig. 2. 

The possible stellar sources of the different groups of SiC grains 
have been discussed elsewhere (e.g., Zinner 1998). Here we wish to 
concentrate on grains of the mainstream 
component (Hoppe et al. 1994; Hoppe \& 
Ott 1997). Their $^{12}$C/$^{13}$C ratios lie between 15 and 100 and their 
$^{14}$N/$^{15}$N ratios between the solar ratio of 272 and 10,000 (Fig. 1). 
Their Si isotopic ratios plot along a line of slope 1.31 in a 
$\delta$$^{29}$Si/$^{28}$Si vs. $\delta$$^{30}$Si/$^{28}$Si 3-isotope 
plot (Fig. 3). Most grains show large $^{26}$Mg excesses attributed to the 
presence of $^{26}$Al ($T_{1/2}$ = 7 $\times 10^5$ yr), now extinct, at
the
time of their formation. Inferred $^{26}$Al/$^{27}$Al ratios range up to 
10$^{-2}$ (Hoppe et al. 1994; Huss et al. 1997). Much more limited isotopic 
data exist for Ti. In Fig. 4 the measurements by Hoppe et al. 
(1994) and Alexander \& Nittler (1999) are plotted as
$\delta$$^{i}$Ti/$^{48}$Ti values against the $\delta$$^{29}$Si/$^{28}$Si 
values of these grains. The correlation between the Ti ratios, especially
$\delta$$^{46}$Ti/$^{48}$Ti, and 
the Si isotopic ratios has already been noticed by Hoppe et al. 
(1994).

There are many pieces of evidence that indicate that mainstream 
SiC grains come from carbon stars. Carbon stars are TP-AGB stars 
whose spectra are dominated by lines of C compounds such as C$_2$, 
CH, and CN, indicating that C $>$ O in their envelopes (Secchi 1868). 
They become C-rich because of the recurrent third
dredge up (TDU) episodes mixing with the envelope newly synthesized   
\cdb from the He shell where it is produced by the triple-$\alpha$ 
reaction  
(Iben \& Renzini 1983). For high-temperature carbonaceous phases, 
such as SiC, to condense from a cooling gas the condition C $>$ O has 
to be satisfied (Larimer \& Bartholomay 1979; Sharp \& Wasserburg 
1995; Lodders \& Fegley 1997a). Carbon stars experience substantial 
mass loss by stellar winds and have extended atmospheres with 
temperatures of 1500 - 2000 K, at which SiC is expected to condense. 
In fact, carbon stars are observed to have circumstellar dust 
shells that show the 11.3 $\mu$m emission feature of SiC (Cohen 1984; 
Little-Marenin 1986; Martin \& Rogers 1987; Speck, Barlow, \& 
Skinner 1997). Recently, Clayton, Liu, \& Dalgarno (1999) proposed
that in a SN environment with high levels of ionizing  $\gamma$-rays carbon 
dust can condense from a gas of O $>$ C. However, even if this should 
be possible, it would not apply to the expanding atmospheres of 
carbon stars. In contrast to such atmospheres, the solar system 
is characterized by O $>$ C and phases such as SiC are not believed 
to be able to form under these conditions. This apparently is the 
reason why all SiC grains found in primitive meteorites are of presolar
origin according to their isotopic compositions. This is in marked 
contrast to presolar corundum grains, which make up only a small fraction
($\sim$1\%) of all meteoritic corundum grains.

Isotopic compositions of presolar grains are the most diagnostic 
indicators of their stellar sources. As already mentioned in 
\S 1, the \s-process patterns of the heavy elements exhibited 
by mainstream SiC grains constitute the most convincing argument 
for their origin in carbon stars, which are the major source of the  
\s-process elements in the Galaxy. The envelopes of these stars show 
large enhancements of typical \s-elements such as Sr, Y, Zr, Ba, La, 
Ce, and Nd (Smith \& Lambert 1990). In SiC grains $s$-process patterns 
are seen in the elements Kr, Xe, Sr, Ba, Nd, Sm, and Dy (Lewis et al. 
1990, 1994; Ott \& Begemann 1990a,b; Prombo et al. 1993; Richter, Ott, 
\& Begemann 1993, 1994; Zinner, Amari, \& Lewis 1991; Podosek et al. 
1999), which have been measured in bulk samples. Although these 
samples were collections of all  SiC grain types, there is little 
doubt that the isotopic results must have been dominated by 
contributions of mainstream grains.

In addition, isotopic analyses of Sr, Zr, and Mo by resonance 
ionization mass spectrometry (RIMS) have been made on a limited 
number of single SiC grains. Although most of these measurements 
were made on SiC grains for which no C and Si isotopic data had 
been obtained (Nicolussi et al. 1997, 1998a,b), for statistical 
reasons essentially all of these grains must have been mainstream 
grains. They display characteristic \s-process patterns with large 
depletions in the $p$-only isotopes $^{84}$Sr, $^{92}$Mo, and 
$^{94}$Mo, and in the $r$-only isotope $^{100}$Mo. Large depletions 
are also seen in $^{96}$Zr, indicating that neutron densities in the 
stellar sources of these grains must have been low, compatible with 
\ctb being the major neutron source in AGB stars (Gallino et al. 
1998a). The depletions in $^{96}$Zr are consistent with isotopic 
abundance data
obtained by spectroscopic observations of the ZrO bandheads in AGB star 
envelopes (Lambert et al. 1995). Recent RIMS analysis of Mo in SiC grains 
that had been identified as mainstream grains on the basis of their 
C, N, and Si isotopic ratios confirmed that mainstream grains 
indeed carry \s-process signatures in this element (Pellin et al. 
1999). Gallino et al. (1997) could successfully reproduce the measured 
\s-process compositions of the heavy elements in mainstream SiC grains
with models of low-mass 
AGB stars of close-to-solar metallicity, in which neutrons are primarily 
produced by the \ctb source during the radiative interpulse period. In
addition to the isotopic patterns of the heavy elements, the large 
overabundances of refractory \s-process elements such as Zr, Y, Ba, 
and Nd in single SiC grains (Amari et al. 1995b) are further evidence 
for an AGB-star origin (Lodders \& Fegley 1995, 1997a,b, 1998).

Evidence for a carbon star origin is also obtained from the light 
elements. It should be noted that as far as neutron-capture 
nucleosynthesis is concerned, because of their large abundance 
the light elements (elements lighter than Fe) in AGB stars are 
considered to be neutron poisons for the synthesis of the 
heavy elements. However, because of their relatively small cross 
sections, they are only marginally affected by neutron capture. 
Elements up to Mg are also affected by charged particle reactions 
with H and He. The most important light-element signature of SiC 
is the Ne isotopic composition, which is dominated by $^{22}$Ne  
(Lewis et al. 1990, 1994). In fact, it has been the presence of 
this Ne component, Ne-E(H), that, together with the so-called 
Xe-S component  (Srinivasan \& Anders 1978; Clayton \& Ward 1978) 
led to the isolation of presolar SiC (Tang \& Anders 1988a). 
Gallino et al. (1990, 1994) showed that Ne-E(H) matches the 
predicted isotopic composition of Ne in the He shell of AGB stars. 
Almost all initial CNO nuclei are first converted to $^{14}$N during 
shell H burning and then to $^{22}$Ne via the chain \nqagb in the 
He shell during thermal pulses. 
Another piece of evidence is obtained from the distribution of the  
$^{12}$C/$^{13}$C ratios in mainstream grains that is very similar to that 
measured astronomically in carbon stars (Dominy \& Wallerstein 
1987; Smith \& Lambert 1990; see also Fig. 14 in Anders \& Zinner 
1993).

The ranges of $^{12}$C/$^{13}$C and $^{14}$N/$^{15}$N ratios 
measured in mainstream grains roughly agree with the ranges predicted 
by theoretical 
models of AGB stars. Proton captures occurring in the deep envelope 
during the main sequence phase followed by first (and second) 
dredge-up as well as shell He burning and the TDU during the 
TP-AGB phase affect the C and N isotopes in the envelope.
$^{12}$C/$^{13}$C ratios
predicted by canonical stellar evolution models range from $\sim$ 20 
at first dredge-up in the RG phase to $\sim$ 300 
in the late TP-AGB phases (Iben 1977a; Bazan 1991; Gallino et al. 
1994). Predicted $^{14}$N/$^{15}$N ratios are 600 - 1,600 (Becker \& 
Iben 1979; El Eid 1994), falling short of the range observed in the grains. 
However, the assumption of deep mixing (``cool bottom processing'' or 
CBP) of envelope material to deep hot regions in $M \lesssim$ 2.5 
\msb stars during their RG and AGB phases (Charbonnel 1995; Wasserburg,  
Boothroyd, \& Sackmann 1995; see also Langer et al. 1999a for 
rotationally induced mixing) results in partial H burning, with
higher $^{14}$N/$^{15}$N and lower $^{12}$C/$^{13}$C ratios in the envelope
than in canonical models (see also 
Huss et al. 1997). 
As a matter of fact, CBP mechanisms have been introduced to explain the
observed 
$^{12}$C/$^{13}$C ratios in RG stars of low mass 
(Gilroy 1989; Gilroy \& Brown 1991;  
Pilachowski et al. 1997), which are lower than those 
predicted by canonical models.

In contrast to the heavy elements, and the light elements C, N,  
Ne and Al, the Si isotopic ratios of mainstream SiC grains cannot 
be explained by nuclear processes taking place in a single star. 
In Fig. 3 we plotted the Si isotopic data measured in SiC grains 
from three different size fractions isolated from the Murchison 
carbonaceous (CM2) meteorite (Hoppe et al. 1994, 1996a) and data 
from the Orgueil (CI) meteorite (Huss et al. 1997). The three 
Murchison size fractions are KJE (0.5 - 0.8 $\mu$m in diameter), 
KJG (1.5 - 3 $\mu$m), and KJH (3 - 5 $\mu$m) (Amari et al. 1994). 
Of the smallest Murchison grain size fraction KJE we plotted only 
data points with errors smaller than 15 \permil. The distributions of 
Si isotopic ratios measured in SiC grains from other meteorites 
are very similar to that shown in Fig. 3 (Alexander 1993; Huss, 
Fahey, \& Wasserburg 1995;  Gao et al. 1995). Also plotted in 
Fig. 3 is the correlation line obtained from a fit to the grain 
data. This line, which does not go through the solar composition 
but passes slightly to the right of it, has a slope of 1.31 and 
an intercept of the ordinate at $\delta$$^{29}$Si/$^{28}$Si$_{int}$ 
= $-$ 15.9 \permil, a little different from the parameters determined from 
the KJG and KJH dataset only (Hoppe et al. 1994).

There have been various attempts to explain the Si isotopic 
distribution of the mainstream SiC grains. Zinner et al.  (1989) 
already realized that the scatter in the Si isotopic ratios 
indicate several stellar sources. Stone et al. (1991) first 
noticed the correlation line of Si isotopic ratios in SiC grains 
from Orgueil and proposed an origin in a single AGB star 
with mixing of two components but they did not address the question of 
how the end components could be generated by nucleosynthetic processes 
in a single star. The only nuclear reactions in AGB stars that are 
believed to substantially affect the Si isotopes are neutron captures 
in the He shell. However, it has been determined early on (Gallino et 
al. 1990; Obradovic et al. 1991; Brown \& Clayton 1992b), and will be 
seen in more detail in the \S 3, that neutron captures shift the 
Si isotopic ratios along a line with a slope that varies, depending 
on mass and metallicity, from 0.35 to 0.75 in a 
$\delta$$^{29}$Si/$^{28}$Si vs. $\delta$$^{30}$Si/$^{28}$Si 
3-isotope plot. Furthermore, in low-mass AGB stars of 
close-to-solar metallicity predicted shifts of envelope material 
are only on the order of 20 \permil, an order of magnitude less than 
the range seen in mainstream grains. Nuclear processes in a single 
AGB star therefore cannot produce the mainstream distribution and 
this led to the conclusion that several stars with varying initial 
Si isotopic compositions must have contributed SiC grains to the 
solar system  (Clayton et al. 1991; Alexander 1993). The situation 
is similar for Ti. It should be emphasized that there is a 
fundamental qualitative difference between the isotopic compositions 
of C, N, Ne, and the heavy elements, and those of Si and Ti. While 
the former are dominated by RG and AGB nucleosynthesis, the effect 
of stellar nucleosynthesis on the Si and Ti ratios is relatively 
small and cannot explain the compositions observed in grains; the 
presence of an extra component has to be invoked.

Brown \& Clayton (1992b, 1993) proposed a single-star model by 
considering Mg burning at elevated temperatures in the He-burning 
shell. In this model ($\alpha$,n) reactions on Mg in a 5.5 \msb 
AGB star produce a neutron-rich Si isotopic composition at the far 
end of the mainstream correlation line. Mixing with the original, 
close to solar, composition in the envelope combined with variable 
mass loss from this star could lead to the observed distribution. 
However, to accomplish this, the temperature in the He shell 
has to be raised by 10\% above that produced by the standard AGB 
models (Iben 1977b). This leads to serious problems with other 
processes and with the general question of energy generation and 
stellar structure. Moreover, the Ti isotopic variations and 
especially the correlation with the Si isotopic ratios (Fig. 4) 
cannot be explained in this way because the temperatures required 
for Mg burning in the He shell do not affect at all the Ti 
isotopes, nor would the \s-process isotopic signatures be compatible 
with such a situation.

This leaves variations in the initial Si isotopic compositions of 
the AGB stars that contributed SiC grains to the solar system as 
the most likely explanation for the Si isotope distribution of 
mainstream grains. Variations of the initial Si compositions in 
turn are expected as the result of GCE. Low-mass stars, 
which became AGB stars at the end of their evolution and 
contributed grains to the protosolar nebula, are likely to 
have been born at different times before solar system formation 
and thus reflect the isotopic composition of the Galaxy in 
different earlier epochs. An explanation of the Si isotopic 
compositions in mainstream grains thus requires an understanding 
of the evolution of the Si isotopic ratios throughout Galactic 
history.

Gallino et al. (1994) approached this problem by assuming that 
$^{29}$Si and $^{30}$Si are mostly primary isotopes that are 
produced by SNe of Type II, together with a major fraction of 
$^{28}$Si, and that substantial contributions to Galactic Si 
in the form of almost pure $^{28}$Si from SNe of Type Ia late 
in Galactic history determine the Si isotopic evolution reflected 
by the grains. 
These authors also advanced a tentative interpretation of the 
Ti isotopes. They noticed the correlation between the Ti and 
Si isotopic compositions (Fig. 4) and concluded that, as for 
Si, neutron-capture nucleosynthesis cannot explain the Ti 
isotopes nor the correlation with Si and they would have to be 
interpreted within the framework of the chemical evolution of 
the Galaxy.

Timmes \& Clayton (1996) and Clayton \& 
Timmes (1997a,b) constructed a detailed model of the Galactic 
history of the Si isotopes that is based on the GCE model of Timmes, Woosley 
\& Weaver (1995). The SN production yields were obtained from 
the Type II SN models of Woosley \& Weaver (1995, henceforth WW95)  
and from the popular W7 Type Ia SN model by Thielemann, Nomoto, \& 
Yokoi (1986). According to the WW95 models, $^{29}$Si and 
$^{30}$Si in Galactic disk stars are predominantly 
secondary isotopes, i.e. their production in massive Type II SNe 
increases with increasing metallicity, since it 
requires the prior presence of primary isotopes such as $^{12}$C,
$^{14}$N and $^{16}$O.
As a consequence, early SNIIe produced mostly pure $^{28}$Si, 
whereas later SNIIe added more and more $^{29}$Si and $^{30}$Si 
to the ISM. This resulted in a continuous increase of the 
$^{29}$Si/$^{28}$Si and $^{30}$Si/$^{28}$Si ratios throughout 
Galactic history and the distribution of the grains' Si isotopic ratios 
apparently reflects this change. According to the Timmes \& Clayton 
model, the spread in the isotopic compositions of the mainstream grains 
corresponds to variations in the birth dates of their parent stars. 
Specifically, the birth dates of the parents stars range over $\sim$ 
5 Gyr (see Fig. 6 in Timmes \& Clayton 1996).

However, this model suffers from a fundamental problem. Because 
most mainstream grains have $^{29}$Si/$^{28}$Si and 
$^{30}$Si/$^{28}$Si ratios that 
are larger than those of the solar system (Fig. 3), they are 
inferred to be younger than the sun. This absurd corollary of 
the model led Clayton (1997) to consider the possibility that 
the mainstream grains originated from stars that were born at 
different Galactic radii than the sun. Indeed, according to 
Wielen, Fuchs, \& Dettbarn (1996), scattering 
by massive molecular clouds may lead to the diffusion of those stars 
from central metal-rich regions of the Galaxy (for which higher 
$^{29}$Si/$^{28}$Si and $^{30}$Si/$^{28}$Si ratios are predicted 
than those in the present solar neighborhood) to the region where 
they, once they became AGB stars, shed their SiC grains into the 
protosolar cloud. 
Alexander \& Nittler (1999) took a different approach to resolve the 
paradox that the grains are apparently younger than the sun. They 
fitted the Si and Ti isotopic compositions of the mainstream grains 
to contributions from nucleosynthesis in the parent AGB stars and 
the stars' original isotopic compositions. However, in contrast to 
the Timmes \& Clayton (1996) model, they tried to determine the 
evolution of the Si and Ti isotopes from the grain data themselves. 
From this fit they concluded that most mainstream grains 
do not come from stars with higher than solar metallicities but that 
the sun has an atypical Si isotopic composition.

Common to the GCE models listed above is that they predict a 
monotonic relationship between the Si isotopic composition and 
Galactic time for the ISM at a given Galactic radius. In other 
words, a given age of the Galaxy or, if different Galactic radii 
are considered, a given function of age and Galactic radius 
corresponds to a given isotopic composition (this functional 
relationship is implied in the Clayton \& Timmes (1997a) model 
but has never been explicitly worked out).

However, one has to consider that the Si isotopic composition 
of the Galaxy, and in particular that of different regions where 
individual low-mass stars are born, evolves as the result of 
contributions from discrete stellar sources, mostly SNe. It 
is unlikely that these discrete contributions are instantly 
mixed with preexisting material so that the isotopic compositions 
of large Galactic regions are completely homogenized. The work 
of Edvardsson et al. (1993) and others has shown that stars 
from any given Galactic epoch and Galactic radius display a 
considerable spread in metallicity and, more generally, in elemental abundances. 
Numerous attempts have been made to explain these observations: 
stellar orbital diffusion, chemical condensation processes 
and/or thermal diffusion in stellar atmospheres, incomplete mixing of 
stellar ejecta, sequential stellar enrichment and local infall of 
metal-deficient gas (see van den Hoek \& de Jong 1997 for review and 
discussion of these different attempts).

The study of meteorites provided ample evidence that the protosolar 
nebula was isotopically not homogenized. In addition to the survival 
of pristine stardust, isotopic anomalies are found in material that 
apparently was processed in the solar nebula (see, e.g., Clayton, 
Hinton, \& Davis 1988; Lee 1988; Wasserburg 1987). Examples include 
$^{16}$O excesses (up to 5\%) and large deficits and excesses of the 
n-rich isotopes of the Fe-peak elements such as $^{48}$Ca and $^{50}$Ti 
in refractory inclusions. These anomalies indicate the survival of 
isotopic signatures from different nucleosynthetic reservoirs. Another 
indication of local isotopic heterogeneity comes from the value of the 
$^{17}$O/$^{18}$O ratio, which is 5.26 in the solar system but 
3.65 $\pm$ 0.15 in diverse molecular clouds (Penzias 1981; Wannier 
1989; Henkel \& Mauersberg 1993, see also discussion concerning 
the $^{14}$N/$^{15}$N ratio by Chin et al. 1999).

We therefore want to explore to what extent the Si isotopic 
spread of mainstream SiC grains can be explained by local 
heterogeneities in the regions from which the low-mass parent 
stars of the grains originally formed. Before doing this we 
will examine in more detail the nucleosynthesis of the Si 
and Ti isotopes in AGB stars. However, because of the largely 
uncertain astrophysical origin of all Ti isotopes  (see  
Timmes et al. 1995; Woosley 1996; Woosley et al. 1997), the 
situation concerning the Ti isotopes in presolar SiC 
grains and their interpretation in terms of SN contributions 
and GCE is a complicated issue in itself. It will be treated 
in a separate paper.

\section{Nucleosynthesis of the Si and Ti isotopes in AGB stars}

During all the evolutionary phases of low-mass stars ($M <$ 10 \ms) the
maximum temperature in the inner regions never reaches high enough values
to allow the burning of any element heavier than He. 
Consequently, only the production of \cdb and of $^{16}$O from
initial H or He nuclei is possible and, in particular, there are no 
charged-particle interactions that involve the nucleosynthesis of 
Si and Ti.
The initial isotopic compositions of these elements can be nevertheless  
modified by slow neutron capture (the $s$ process), which occurs
in the tiny region between the H shell and the He shell (hereafter He
intershell) during the AGB phase. 

According to the AGB models of low-mass ($M =$ 1.5 $-$ 3 \ms) stars with 
metallicities in the range from half solar to solar obtained with the FRANEC
code and discussed in detail 
by Straniero et al. (1997) and Gallino et al. (1998a), neutrons are 
released in the He intershell by two different sources, \ctb and 
$^{22}$Ne. The maximum temperature achieved during the 
recurrent thermal instabilities (or thermal pulses: TP) of the He 
shell is not high enough to consume $^{22}$Ne to an appreciable extent, so 
that the \ctb source has to play the major role. However, 
the number of \ctb nuclei left behind by the H-burning shell is too 
small to account for the \s-element enhancements observed in carbon 
stars. A special mechanism has to be invoked in order to build up a 
sufficient amount of \ctb in the He intershell. In AGB stars of mass 
$M \ga 1.5$ \ms, after a limited number 
of thermal pulses, soon after the quenching of a given instability, 
the convective envelope penetrates into the top layers of
the He intershell and mixes with the envelope  
material enriched in \cdb and $s$-process elements. This TDU 
phenomenon leaves a 
sharp H/He discontinuity, where some kind of hydrodynamical mixing, 
possibly driven by rotation, occurs 
(Herwig et al. 1997; Singh, Roxburgh, \& Chan 
1998; Langer et al. 1999a,b). In these conditions, a small 
amount of protons penetrates from the 
envelope into the He intershell (see Gallino et al. 1998a for 
discussion). At H reignition, these protons are captured by the 
abundant \cdb present in the intershell as 
a consequence of partial He burning that occurred during the previous 
thermal instability. Consequently, a so-called $^{13}$C {\it pocket} 
is formed in a small region at the top of the 
He intershell. Before the onset of the next pulse, the progressive 
compression and heating of these layers cause all \ctb to burn
radiatively in the interpulse period 
via the \ctanb reaction, at a temperature of around 8 keV. 
The neutron exposure, or time integrated neutron flux 
$\delta \tau = \int {\it N}_{\rm n}$ $v_{th}~ dt$, experienced
during the interpulse period may reach quite a 
high value, $\delta \tau_1 (8  {\rm keV})$ of up to 0.4 mbarn$^{-1}$, 
depending on the initial amount of \ctb. The maximum neutron density in this 
radiative phase remains low: $N_{{\rm n}, max}$ $\sim$ 10$^7$ n/cm$^{3}$.

The material that experienced neutron captures in the \ctb pocket 
is engulfed and diluted ($\sim$1/20) 
by the next growing convective thermal pulse, which extends over almost 
the whole He intershell, and is mixed with material already $s$-processed 
during the previous pulses, together with ashes of the H-burning 
shell. Among them are Si and Ti in their initial abundances. For advanced 
pulses, the overlapping factor between subsequent pulses becomes $r$ 
$\approx$ 0.4 
and the mass of the convective pulse is slightly smaller than 10$^{-2}$ \msb 
(Gallino et al. 1998a). A small neutron burst, at around 23 keV, is
released in convective thermal pulses, during the latest phases of the
AGB evolution, when the bottom temperature in the He shell is sufficiently
high to marginally activate the \neanb reaction. The $^{22}$Ne is provided by
the 
\nqagb chain starting from $^{14}$N present in the ashes of H burning.
Some extra $^{14}$N derives from primary $^{12}$C that is dredged up into the
envelope and partly converted to $^{14}$N by H-shell burning. The neutron
exposure provided by the $^{22}$Ne neutron source during the thermal 
pulse is low, reaching at most
$\delta \tau_2 (23 {\rm keV}) =$ 0.03 mbarn$^{-1}$.
However, the peak neutron density can reach 10$^{10}$ n/cm$^{3}$.

Exposure to the two neutron fluxes is repeated through the 
pulses with TDU. 
The \ct-pocket features are kept constant pulse after pulse, 
while the small neutron exposure from the $^{22}$Ne neutron source 
during thermal pulses increases 
with pulse number, reflecting the slight increase of the strength of the 
thermal instability with core mass.

We have performed new calculations for the nucleosynthesis
due to neutron capture in AGB stars. 
These calculations are based on the
stellar models and the nuclear network described in detail by Gallino
et al. (1998a). We want to follow here in particular the modifications of 
the Si and Ti isotope abundances arising by neutron capture 
in the intershell region during the neutron fluences in the 
\ctb pocket and in the thermal pulses. Then we will follow the isotopic 
compositions of Si and Ti in the envelope as they are modified 
during the whole TP-AGB phase by the mixing of He intershell material 
due to TDU episodes. The envelope itself is progressively eroded by stellar
winds and by the growth of the H-burning shell.

In Table 1 the Maxwellian averaged neutron capture cross sections (in the
form of $\sigma_{code}$=$\left< \sigma v \right>/v_{th}(30 {\rm
keV})$, expressed in mbarn) are listed for selected isotopes. 
Values are reported for the two typical temperatures 
(8 keV and 23 keV) at which neutrons are released by the \ctb and by the
\neonb neutron source, and for the standard temperature of 30 keV at 
which these cross sections are currently given in the literature. The
quoted values have been 
taken from the compilation by Beer, Voss, \& Winters (1992), with three 
exceptions: $^{28}$Si, for which a renormalization to the 30 keV 
recommended value of Bao \& K\"appeler (1987) has been considered 
(Beer 1992, private communication), $^{150}$Sm, from Wisshak et al. (1993), 
and the $^{33}$S(n,$\alpha$)$^{30}$Si cross section from Schatz et al. 
(1995). The use of $\sigma_{code}$ demonstrates how the cross section
at any given energy departs from the usual $1/v$ rule. For a perfect 
$1/v$ dependence, $\sigma_{code}$ at different $kT$ values should remain
constant. In reality, with the exception of $^{49}$Ti, strong 
departures from this rule are shown by all Si 
and Ti isotopes. The $^{30}$Si abundance 
resulting from a given neutron exposure is strongly dependent on the
reaction rate of $^{32}$S(n,$\gamma$)$^{33}$S, where $^{33}$S is subsequently quickly transformed to $^{30}$Si via $^{33}$S(n,$\alpha$)$^{30}$Si.

The neutron capture cross sections of two typical 
heavy $s$-only nuclei, $^{100}$Ru and $^{150}$Sm, have also been included
in Table 1  
in order to show that Si and Ti (this is also true for all 
other elements lighter that Fe) are not as much affected by neutron 
capture as the heavier elements. The neutron capture cross sections of 
light elements are much smaller (by as much as three 
orders of magnitude) than 
those of typical heavy isotopes. However, because of their large 
initial abundances, isotopes lighter than $^{56}$Fe act as important 
neutron poisons for the build-up of the heavy elements. 
In the last column, the relative uncertainties of the 30 keV cross sections are 
reported. The cross sections of all Si and Ti isotopes still 
suffer from large uncertainties, around 10\%, 
whereas for many heavy isotopes recent experiments have achieved a 
precision of the order of 1\% (see K\"appeler 1999). 

Table 2 shows the production factors with respect to solar 
of the Si and Ti isotopes, as well as that of two 
\s-only nuclei $^{100}$Ru and $^{150}$Sm, 
in the He intershell at different phases of the 15$^{\rm th}$ thermal 
pulse for an AGB star of 1.5 \msb and solar metallicity, with the 
standard choice of the \ctb pocket (case ST of Gallino et al 1998a). 
Columns 2 and 3 show the effect of the \ctb neutron source on the Si 
and Ti isotopic abundances inside the \ctb pocket. The values of
column 4 were calculated at the time when the $s$-enriched pocket
has been engulfed by the growing convective pulse and diluted
with both $s$-processed material from
the previous pulse and material from the H-burning ashes, containing in
particular Si and Ti of initial composition. 
The difference between the values of columns 4 and 5 
expresses the effect of the $^{22}$Ne neutron source activated during the
15$^{\rm th}$ pulse. Note how $^{100}$Ru and $^{150}$Sm production factors are up 
to three orders of magnitude larger than those of Si and Ti. 
From the results given in Table 2, it is easily recognized that 
neutron captures only marginally modify the initial Si and Ti 
isotopic compositions. $^{28}$Si tends to be slightly consumed (by 
15 \permil, and by 20 \permil, respectively) after both neutron exposures. 
Actually, during 
the high neutron exposure from the \ctb source $^{29}$Si is more 
efficiently consumed (by a factor 1.6) than produced, because of the 
very low cross section of $^{28}$Si. Also $^{30}$Si is consumed (by 
a factor 2.2) during this phase. In contrast, the abundances of both 
neutron-rich Si isotopes grow during the thermal pulse, by factors
of about 1.3 and 1.4, respectively, relative 
to their initial values in the convective 
pulse. These features are mainly due to the fact that the neutron 
capture cross sections strongly depart from the  1/$v$ trend. Note 
that $\sigma_{code}(^{28}$Si) is almost an order of magnitude greater at
23 keV than at 8 keV, which explains why in the TP phase this isotope is 
destroyed to a larger extent than in the \ctb pocket. 
As a consequence, we observe the 
growth of $^{29}$Si during the TP. 

Among the Ti isotopes, $^{50}$Ti is a neutron magic nucleus ($N$ $=$
28) and its neutron capture cross section is very small compared to 
those of the other Ti isotopes. It shows a strong departure from the 
1/$v$ trend (see Table 1), being a factor of 4 greater at 23 keV than 
at 8 keV. As shown in Table 2, $^{50}$Ti accumulates during the 
\ctb neutron exposure because of its very low neutron capture cross 
section, which makes this isotope a bottleneck of the abundance flow. 
$^{49}$Ti is produced in both phases, by a larger factor during the 
pulse, while $^{48}$Ti is consumed. Note that during the high neutron 
exposure by the \ctb neutron source $^{48}$Ti is only marginally 
modified, despite its relatively large cross section. This results 
from abundance flow starting at $^{40}$Ca, an isotope of large initial 
abundance and of cross section $\sigma_{code}$(8 keV) $=$ 6.11 mbarn, 
which is consequently consumed (by a factor $\approx$ 6) in this phase. 
The $^{46}$Ti, $^{47}$Ti and $^{48}$Ti isotopes, similarly to the 
Si isotopes, suffer almost negligible variations. 

Because of their relatively small cross sections, a behavior similar to 
that of Si and Ti is shown by other light elements below Fe, among 
them S and Ca. It should be emphasized that the final 
Si isotope composition mostly depends on the small neutron exposure by 
the $^{22}$Ne neutron source in the convective pulse rather than on 
the very large neutron exposure by the \ctb neutron source taking 
place in the tiny radiative \ctb pocket.  

Table 2, column 6 shows the production factors in the envelope 
immediately after the TDU that follows the
quenching of the 15$^{\rm th}$ thermal pulse. At this stage, the star has
become a C star, with C/O $=$ 1.3, and the isotopic composition of the
envelope results from the mixing of the $s$-processed and \cd-enriched 
material cumulatively carried into the envelope by previous TDU episodes.

Predictions for the Si and Ti isotopic compositions in the envelope of AGB
stars of solar metallicity and initial mass of 1.5 and 3 \msb during
repeated TDUs in the TP phase are shown in 
Fig. 5. Fig. 6 reports predictions for the resulting Ti vs. Si correlation:
as in Fig. 4, Ti ratios are plotted as function of the
$^{29}$Si/$^{28}$Si ratio. They are all reported in the form of 
$\delta$-values for three different choices of the amount of \ctb in the 
He intershell. The standard case (ST) of Gallino et al. (1998a) 
corresponds to an average mass fraction of $^{13}$C of 6 $\times$ 
10$^{-3}$ distributed over a tiny layer of a few 10$^{-4}$ \msb at the top 
of the He intershell, case d3 corresponds to the amount of case ST 
divided by 3 and case u2 is an upper limit corresponding to the amount 
of case ST multiplied by 2. As already mentioned in Busso et al. (1999a) 
(see also Busso, Gallino, \& Wasserburg 1999b), a spread in the $^{13}$C 
amount in stars of different metallicities
is required by spectroscopic observations of  
$s$-enhanced stars, and conceivably depends on the initial stellar mass or 
other physical characteristics (such as stellar rotation).
The measurements of Zr, Mo, and Sr isotopic ratios in individual SiC 
grains have confirmed this spread in the \ctb amount: all single 
grain compositions can be matched by low-mass AGB models of about solar 
metallicity if we consider different amounts of \ctb (Gallino et al. 
1998b; Nicolussi et al. 1998b). Open symbols are for 
envelopes with C/O$>$1, the condition for SiC condensation. 
Note that case ST for solar metallicity best reproduces the $s$-process
isotopic distribution of bulk SiC grains, which is slightly different from
the solar main component (for a general discussion see Gallino et al. 1997; 
Busso et al. 1999b).

Not surprisingly the $^{29}$Si/$^{28}$Si, $^{30}$Si/$^{28}$Si, and 
$^{47}$Ti/$^{48}$Ti ratios are only a few percent (up to 25 \permil, 
40 \permil, and 
14 \permil, respectively) higher than the corresponding solar ratios. The 
$^{46}$Ti/$^{48}$Ti and $^{49}$Ti/$^{48}$Ti ratios are 
up to 70 \permilb and 200 \permilb higher than the solar ratios. 
In agreement with the results shown in Table 2, the
only ratio that is affected to a significant extent (up to 500 \permilb
higher than solar)
is the $^{50}$Ti/$^{48}$Ti ratio. Note that the 
$\delta^{50}$Ti$/^{48}$Ti values range from $+$100 \permilb to $+$500 
\permil, depending on the \ctb amount. $^{50}$Ti is a
magic nucleus whose abundance is very sensitive to 
the high neutron exposure in the \ctb pocket. 
The fact that the $^{50}$Ti/$^{48}$Ti ratio is significantly changed during 
the AGB phase is, in a way, consistent with the $^{50}$Ti/$^{48}$Ti ratios 
measured in SiC grains. Model predictions do not reproduce the spread of the 
measured Si and Ti compositions, nor could they ever explain the 
negative $\delta$-values measured in some grains; 
the calculated $^{50}$Ti/$^{48}$Ti ratio, 
though, reaches $\delta$-values that are higher than those of 
all the other Si and Ti $\delta$-values 
both in AGB model predictions and in single 
SiC grain measurements (up to 300 \permil, 
see Fig. 4). 

We also investigated two other TP-AGB models: a case of $M=5$ \msb of
solar metallicity (Fig. 7) and a case of $M=3$ \msb of 1/3 solar
metallicity (Fig. 8). An interesting feature is common to both models:
the maximum temperature at the bottom of the He convective shell is
somewhat higher than in the models described above. As a consequence, 
the production factors for $^{29}$Si and
$^{30}$Si, whose production is most sensitive to the \neonb neutron 
source (see Table 2), at the end of the 15$^{\rm th}$ pulse reach 3.2 and 
5.9, respectively, for the $M=5$ \msb star of solar metallicity, and 3.1 
and 6.5, 
respectively, for the $M$ $=$ 3 \msb star of $Z$ $=$ $Z_{\odot}$/3. 

This results in an increase of up to 80 \permilb and 200 \permilb in  
$\delta^{29}$Si$/^{28}$Si and $\delta^{30}$Si$/^{28}$Si, 
respectively, in the envelope of the $M=5$ \msb model (Fig. 7), 
and of up to 100 \permilb and 200 \permilb in the 
1/3 $Z_{\odot}$ model (Fig. 8). The largest $^{29}$Si and $^{30}$Si excesses
measured in SiC are reproduced, however with a slope of about 0.5
for the mixing line, whereas the slope of the mainstream correlation line in 
the Si 3-isotope plot is 1.31 (Fig. 4). 
Note the extremely high values 
(up to 2000 \permil) reached by $\delta^{50}$Ti$/^{48}$Ti in the last 
case (Fig. 8). They
result from the fact that, in our AGB model, the neutron exposure in the
\ctb pocket is very sensitive to metallicity: it grows with
decreasing metallicity (see Gallino et al. 1999).

As for all the cases above, the initial isotopic composition of the star 
has been assumed to be solar, including  the $Z = Z_{\odot}/3$ case. 
In principle, some enhancement for 
isotopes produced by $\alpha$ captures (such as $^{16}$O, $^{20}$Ne, 
$^{24}$Mg, $^{28}$Si, $^{40}$Ca and $^{48}$Ti) as well as complex 
secondary-like trends of many other nuclei should be taken into account 
in the initial composition of low-metallicity stars. This is a tricky point, 
for these variations have to be deduced from GCE models together with 
spectroscopic observations and are, in many cases, not well defined. 
An exercise of this kind, in connection with a possible explanation for 
the Si isotopic composition of SiC of type Z, can be found in 
Hoppe et al. (1997). For the AGB model of
$Z$ $=$ $Z_{\odot}$/3, we made some tests by assuming a small enhancement of the 
initial $^{28}$Si and $^{32}$S, as well as of other $\alpha$-rich 
isotopes according to the spectroscopic evidence by
Edvardsson et al. (1993), and  small depletions in the initial
abundance of  the secondary-like isotopes $^{29,30}$Si. It turned out that
the resulting Si isotope composition in the He intershell as a consequence 
of neutron captures was quite insensitive to the above variations, being
dominated by the most abundant $^{28}$Si. 

It has to be remarked here that several features of the predicted Si and
Ti ratios (e.g., the slope in the Si 3-isotope plot) depend on the
neutron capture cross sections which, for the Si as well as the
Ti isotopes, are still quite uncertain, as shown in
the last column of Table 1. New measurements are highly desirable for 
obtaining the best possible AGB model predictions. 

\section{The stellar sources of Si and Galactic heterogeneity}

\subsection{Supernova sources}

As Timmes \& Clayton (1996) have pointed out, SNe of Type II are 
the dominant sources of Si in the Galaxy, especially in its early 
stages. At later Galactic times, SNe of Type Ia also contribute $^{28}$Si. 
The SNII models of WW95 show that $^{28}$Si is a primary isotope 
whereas $^{29}$Si and $^{30}$Si are predominantly secondary isotopes (see also 
Timmes \& Clayton 1996 for details). This means that $^{28}$Si 
can be synthesized in early Type II SNe from a pure H and He 
composition, whereas the production of $^{29}$Si and $^{30}$Si 
requires the prior presence of primary isotopes such as $^{12}$C, 
$^{14}$N and $^{16}$O. As a 
consequence, the $^{29,30}$Si/$^{28}$Si ratios of the ejecta of 
SNIIe increase with the metallicity of the stars. While $^{28}$Si 
is the product of explosive O burning, both $^{29}$Si and $^{30}$Si 
are synthesized in a narrow region by explosive Ne burning. 
Actually, $^{29}$Si production is restricted to the outer region 
of the Ne burning shell.

Fig. 9 shows the $^{29,30}$Si/$^{28}$Si ratios (plotted as 
$\delta$-values) of the averages of the yields of SNII models of 
different metallicities. The cases of metallicity 
\Z  = 0.1 $Z_{\odot}$ and \Z = $Z_{\odot}$ are taken from 
WW95, the cases \Z = 0.5 $Z_{\odot}$ and 
\Z = 2 $Z_{\odot}$ are from more recent, unpublished 
calculations by Weaver \& Woosley. To obtain the averages we took the 
initial mass function for massive stars into account by weighing the 
contributions from SNIIe of different masses according to $M^{-2.35}$ 
per unit mass interval, the Salpeter initial mass function. Fig. 9
shows that in the WW95 models there exists a fairly good linear 
relationship between the $^{29,30}$Si/$^{28}$Si ratios of the 
average SN yields and the metallicity, demonstrating the secondary 
nature of the heavy Si isotopes. 
The GCE of the Si isotopes is thus believed to have progressed from small 
$^{29,30}$Si/$^{28}$Si ratios at early Galactic times to larger and 
larger ratios as the metallicity of the whole Galaxy increased and 
SNIIe of increasing metallicity contributed their Si to the ISM 
(Timmes \& Clayton 1996).

This process is expected to have resulted in the Si isotopic ratios at the 
time and place of solar formation. However, closer inspection of Fig. 
9 and especially Fig. 10, where the $\delta$-values of the
$^{29,30}$Si/$^{28}$Si ratios of the averages of SNII yields are 
plotted in a Si 3-isotope plot, reveals that the SNII models by Weaver 
\& Woosley do not exactly produce the solar Si isotopic composition. 
It is evident that $^{29}$Si in the presently available 
models is under-produced and the isotopic evolution expected from SNII 
contributions misses the solar isotopic composition (Fig. 10). This 
is a long-recognized problem: Type II SN models under-produce $^{29}$Si 
relative to $^{30}$Si as compared to the solar isotopic ratio (Timmes 
et al. 1995; Timmes \& Clayton 1996; Thielemann, Nomoto, \& Hashimoto 
1996; Nomoto et al. 1997). 
This fact is also demonstrated by a comparison of model predictions and 
the Si isotopic ratios of type X SiC, Si$_3$N$_4$, and low-density 
graphite grains, all of which are believed to originate from Type II SNe  
(Nittler et al. 1995; Travaglio et al. 1999). The Si isotopic ratios 
of these grains have systematically higher $^{29}$Si/$^{30}$Si ratios 
than those predicted by SN models (Zinner et al. 1998;  Travaglio et al. 
1999). In order to achieve the solar ratios, Timmes \& Clayton  (1996) 
proposed multiplying the $^{29}$Si yields of SNII models by a
factor 
of $\sim$ 1.5. We will do likewise in this paper and multiply the 
$^{29}$Si yields by the same factor to obtain the best fit to the 
solar isotopic ratios or to the grain data.

It should be mentioned that there still exist major problems associated 
with the synthesis of the Si isotopes in massive stars. 
As Arnett \& Bazan (1997) pointed out, heterogeneous mixing 
between different layers during the late evolutionary stages 
might have a major effect on the nucleosynthesis of 
certain elements. Bazan \& Arnett (1998) used a two-dimensional 
hydrodynamic code to investigate convective O-shell burning in a 20  
\msb star. They concluded that the results of these calculations differ 
in many ways from those of one-dimensional models and that corresponding 
changes in the nucleosynthesis of Si during this stage are to be 
expected. It remains to be seen whether full nucleosynthetic calculations 
in two- or three-dimensional models can shed light on the problem of 
relative yields of the Si isotopes. Another problem is the relative 
contribution of Type Ia and Type II SNe to the GCE of the heavy elements, 
in particular Fe. Whereas in the Timmes et al. (1995) GCE model Type II 
SNe were assumed to contribute 2/3 of the Fe in the solar system, Woosley 
et al. (1997) favored a more important role of Type Ia SNe, letting them 
contribute as much as half of the solar Fe. This would indicate somewhat 
higher contributions by Type Ia SNe to the Galactic $^{28}$Si relative
to SNIIe. In addition to possible uncertainties in the nuclear physics 
and in the treatment of the various convective zones affecting the 
production of the three Si isotopes, problems are related to the 
effect of mass loss from the most massive stars and to Galactic 
enrichment by close binary massive stars (Woosley, Langer, \& Weaver 
1993, 1995), to the effect of rotation (Heger, Langer, \& Woosley 1999), 
and to the still uncertain development of the explosion (WW95, 
Thielemann et al. 1996).

While in Fig. 9 only averages of SNII models of different metallicities 
have been plotted, it has to be realized that SN models of different masses 
yield very different Si isotopic ratios. In Fig. 10, in addition to
averages, we also plotted the isotopic ratios of individual SNII models 
of different masses for the \Z = 0.1 $Z_{\odot}$ and the 
\Z = $Z_{\odot}$ case. The yields and Si isotopic ratios for the
$Z_{\odot}$ case are also given in Table 3. As can be seen, the isotopic 
ratios of different mass SNIIe span a wide range. There are variations 
not only in the $^{29,30}$Si/$^{28}$Si ratios but also in the
$^{29}$Si/$^{30}$Si ratio. The last column in Table 3 shows the latter 
ratio (already readjusted by augmenting the $^{29}$Si yield) for SNIIe of different
mass. Note that the ratio is smaller than unity for most SNIIe of lower mass 
but larger than unity for the two most massive models. In Fig. 11 we plotted 
again the average Si isotopic ratios for the $Z_{\odot}$ case together with
the averages for the SN models with masses \M $\leq$ 25 \msb and 30 \msb 
$\leq$ \M $\leq$ 40 \ms. This time the theoretical $^{29}$Si yield was
increased 
by the same factor 1.5 for Type II SNe of all masses in such a way that the 
weighted average ratios plot on the slope-one line or, in other words, 
that the average $^{29}$Si/$^{30}$Si ratio is solar. As can be seen, the
low-mass average falls slightly below the slope-one line through the origin 
(pure $^{28}$Si) and the solar isotopic composition and the high-mass 
average falls above this line.

While the evolution of the Si isotopes of the Galaxy as a whole and of 
the average of material in an annulus of a given Galactic radius 
undoubtedly followed the slope-one line, we expect certain variations 
in the Si isotopic ratios even at a given time and a given Galactic radius 
in relatively small regions from which low-mass stars formed. The reason 
is that the addition of contributions from individual SNe, which are 
responsible for the Si isotopic ratios of a certain region, is a stochastic 
process and we do not expect that material from these contributions is 
instantly homogenized with preexisting material. Fluctuations result 
from the fact that individual SN sources have yields with very different 
Si isotopic compositions as is clearly shown in Figs. 10 and 11. In 
addition to the isotopic ratios of Type II SNe, in Figs. 10 and 11 
as well as in Table 3 we show also the ratios of the W7 SNIa model 
(Thielemann et al. 1986; updated by Nomoto et al. 1997) and the SNIa model 
originating from sub-Chandrasekhar (hereafter sub-Ch) white dwarfs accreting 
He from 
a binary companion (Woosley \& Weaver 1994). These SN types are believed 
to be the major sources of Si in the Galaxy around the time of solar 
system formation (Timmes \& Clayton 1996; Woosley et al. 1997). 

Let us now consider the effect on the Si isotopic ratios of the 
admixture of material from one of these SN sources to material with a 
given isotopic composition (Fig. 11). Three-isotope plots such as those 
shown in Figs. 10 and 11 have the property that the isotopic composition 
of a mixture between two components lies on a straight line connecting 
the isotopic ratios of the two components. For the sake of demonstration 
we arbitrarily selected as starting composition the Si isotopic 
composition of the sun. Admixture of different SN sources (we chose  
SNIa W7, SNIa sub-Ch, and, again for the sake of demonstration, the low- and high-mass averages of the 
$Z_{\odot}$ Type II SN models of WW95) will shift the starting 
composition in the directions of the arrows in the figure (see also Figs. 
8 of Timmes \& Clayton 1996 for mixtures between average ISM and 
ejecta from individual SNe. These authors also mentioned the possibility 
of reproducing a larger-than-unity slope for the Si isotopic ratios 
of the mainstream SiC grains but did not systematically develop the 
local heterogeneity picture as it is done in this work). Thus 
admixture of material from a SNIa sub-Ch will shift the 
composition toward the origin (pure $^{28}$Si). We note that the average of
the high-mass SNII sources (with adjusted $^{29}$Si yield) lies above the
slope-one line from the origin through the solar composition. This means 
that admixture of material from these sources will shift, on average, 
the original solar composition along a line with a slope larger than one. 
Likewise, because the average of the low-mass sources lies below the 
slope-one line, admixture from these sources will again, on average, 
result in a shift along a line with a slope larger than one. The same 
is true, even though to a lesser extent, if material from the Type Ia 
W7 SN model (containing essentially pure $^{28}$Si) is added to the solar 
composition.

\subsection{Monte Carlo calculations}

If contributions from a limited number of such sources are 
considered, the resulting Si isotopic compositions will fluctuate from 
one mix to the next because of the statistical nature of these 
contributions. We have developed a simplified Monte Carlo (MC) model in 
which we add material from a limited number of discrete SN sources 
in a statistical way to material with an arbitrary (but reasonable) 
starting isotopic composition in order to see whether the Si isotopic 
distribution of the mainstream SiC grains can be explained as the 
result of statistical fluctuations or, in other words, local 
heterogeneities in the regions where low-mass stars - as AGB stars 
the sources of mainstream grain - were born. A detailed description of 
our MC model can be found in the Appendix.

We have performed different calculations with different assumed starting 
compositions. As first test we took the average Si isotopic ratios of the 
mainstream SiC grains corrected for AGB contributions 
($\delta$$^{29}$Si/$^{28}$Si$_{mean}$ = 30.4 \permilb and
$\delta$$^{30}$Si/$^{28}$Si$_{mean}$ = 27 \permilb - see Appendix) as starting 
composition. In other words, we considered the mean composition of the 
mainstream grains' parent stars as a possible ``standard'' composition 
of the ISM from which these stars were born. We then randomly added \Nb 
SN contributions and randomly chose the sign of each contribution (i.e., 
the sign of the parameter $a$, the constant factor by 
which the total mass ejected by each SN is multiplied) in order 
to simulate material that could 
have seen more or less from each kind of SN source relative to the 
chosen standard ISM composition. In this way we generated 200 different 
mixtures, whose isotopic ratios are plotted Fig. 12a. The plot shows the 
case for \Nb = 100. 
For $a$, the fraction taken from each SN 
source, we obtained $a$ = 1.5 $\times 10^{-5}$ \ms$^{-1}$. Note that
because 
we computed abundances for each isotope $i$ in the form of mass fractions 
$X_i$ (see Appendix), the contributing terms $a \times M_{ejected}$ do 
not have a 
dimension and the parameter $a$ has the dimension of the inverse of a 
mass. As explained in the Appendix, 
because of the limited number (200) of cases, a certain range of the parameter 
\Nb is expected to yield a good fit. Other similarly good matches
are obtained for values of \Nb ranging between $\sim$ 50 and $\sim$ 200, 
and $a$ accordingly from $\sim$ 1.9 $\times 10^{-5}$ \ms$^{-1}$ 
and $\sim$ 0.95 $\times 10^{-5}$ \ms$^{-1}$. In Fig. 12a we also took 
the modification of the Si isotopes by 
nucleosynthesis in the AGB parent stars into account, adding the average 
isotopic shift given above to the results of the Monte Carlo calculation. 
As can be seen from the figure, the 200 different mixtures generated with 
these parameters by MC in a random fashion match the distribution of the 
grains surprisingly well. We note that the slope of the correlation line 
of the MC points is larger than unity but somewhat smaller than the slope 
of 1.31 of the mainstream correlation line. The slope of the MC 
compositions reflects the distribution of the SN sources, mostly the 
SNII sources of solar metallicity. As has been already pointed out above 
in the discussion of Fig. 11, these sources are aligned with an average 
slope that is greater than one.

If we choose starting compositions different from the average of the mainstream 
SiC grains, it turns out that for a wide 
range of starting Si isotopic ratios, as long as they are constrained to 
be compositions expected for the Galactic evolution of the Si isotopes 
(i.e., compositions that in Si 3-isotope plots such as those in Figs. 10 
and 11 lie on the slope-one line between the origin representing pure 
$^{28}$Si and the solar isotopic composition), values for the parameters 
\Nb and $a$ can be found that let us achieve a good match with the Si 
isotope distribution of the mainstream SiC grains, albeit with different 
choices of the parameters \Nb and $a$ for each case. 
We investigated three more cases for which we chose $a$ always to be 
positive. The results are shown in Figs. 12b, 12c, and 12d. In the 
first of these cases the starting isotopic composition is solar, in 
the second case Si is depleted in the heavy isotopes by 100 \permilb 
($\delta$$^{29}$Si/$^{28}$Si$_{init}$ = $-$ 100 \permilb and
$\delta$$^{30}$Si/$^{28}$Si$_{init}$ = $-$ 100 \permil) and 
in the third by 200 \permil. The best-fit parameters are \Nb = 70 and 
$a$ = 1.7 $\times 10^{-5}$ \ms$^{-1}$ for the first case, 
\Nb = 420 and $a$ = 1.1 $\times 10^{-5}$ \ms$^{-1}$ for the second case,
and 
\Nb = 600 and $a$ = 1.5 $\times 10^{-5}$ \ms$^{-1}$ 
for the third case. Also in these cases, we obtain good fits for a range 
of \Nb and $a$ values.

It is clear that the addition of SN material will change the concentration 
of other elements as well. As the statistical nature of these additions 
results in a range of Si isotopic ratios, we expect it to result in a 
corresponding range of elemental ratios as well and these variations can 
be compared with astronomical observations in stars. In Fig. 13a we
plotted the scatter in elemental ratios obtained by the MC calculation 
for the case with $\delta$$^{29}$Si/$^{28}$Si$_{init}$ =
$\delta$$^{30}$Si/$^{28}$Si$_{init}$ = 0. In different Galactic 
regions elemental ratios relative to H are expected to be affected by newly 
infalling gas and not only by the contributions from stellar nucleosynthesis. 
For this reason we plotted ratios relative to Fe and normalized to the
solar system abundances (i.e., [Elem/Fe] = 
log[(Elem/Fe)/(Elem/Fe)$_{\odot}$]). The spreads in the theoretical 
elemental ratios are quite modest, especially if compared with ratios 
observed in stars. Edvardsson et al. (1993) measured elemental abundances 
in a large number of stars from the Galaxy and concluded that stars 
from a given epoch (i.e. of a given age) and from a given Galactic 
radius show a considerable spread in metallicity. That this spread is 
not simply the result of variations in the amount of newly infalling 
material is shown by the fact that also elemental ratios between elements, 
in particular relative to Fe, show considerable variations (Fig. 13b). 
Edvardsson et al. (1993) pointed out that, despite observational errors 
(including a typical uncertainty of 1 - 2 Gyr in the age of dwarf stars), 
these variations are intrinsic (see also Timmes et al. 1995). 
A comparison of Figs. 13a and 13b shows that the spread in elemental 
ratios obtained from a model of statistical fluctuations in the 
contributions from various SN sources is smaller than the spread 
observed in stars. Since we do not know exactly how much of the spread 
in the Edvardsson et al. (1993) data is due to experimental errors, we 
just want to emphasize that the MC spread is not larger than that in 
stars. The models by Copi (1997) and van den Hoek \& de Jong (1997), which 
make use of stochastic approaches in the study of GCE, are able to 
account for these elemental spreads. The mass of a well-mixed 
region (a sort of mixing scale) that in the Copi (1997) model 
yields a good fit to the spread of the abundances 
of $\alpha$-elements (such as Si) in stars 
is $M \sim 10^5$ \ms. If we interpret our constant 
parameter $a$ as the inverse of the total mass of the region in which the 
SN ejecta are expected to be well mixed, we find for this region a mass of $M 
= 1/a \sim 10^5$ \ms, a number remarkably similar to that found by Copi (1997).

We conclude that local heterogeneities in Galactic regions that 
can explain the variations in Si isotopic ratios observed in the 
mainstream SiC grains imply variations in elemental ratios that are 
compatible with those observed in stars. In principle, 
such heterogeneities could be the cause of the mainstream isotopic 
variations.

\subsection{Discussion}

It has to be emphasized that our model for explaining the Si isotopic
variations in mainstream SiC grains does not pretend to fully
simulate the isotopic compositions of the SiC mainstream grains. 
It uses theoretical yields of the Si isotopes ejected from SNe that 
needed adjustment to explain the composition of the solar system (see
section 4.1). In addition, the model is overly simplistic and at 
this point should only be understood as a demonstration that local 
heterogeneities due to the statistical nature of SN contributions 
can in principle successfully reproduce these variations. In reality 
the situation is expected to be much more complicated:

\begin{itemize}

\item {1) There will be a statistical spread in the individual SN 
contributions;}
\item {2) There will be a spread in the initial composition;} 
\item {3) There will be contributions from SNe with a range of 
metallicities;} 
\item {4) There will be a range in ages of the AGB stars because 
of differences in their mass.}

\end{itemize}

We will discuss these points in turn. 

\begin{itemize}

\item {1) We assumed that all the SN sources that add material to a given 
Galactic region contribute the same amount as expressed by a single 
value of the parameter $a$. In reality different SNe will contribute 
different amounts and in some extreme cases one SN will completely 
dominate the local mix. In our MC calculations we have also assumed 
different statistical distributions for the parameter $a$ and could 
achieve essentially the same final results as those shown in Figs. 12.}

\item {2) We have shown that different initial Si isotopic compositions can 
produce distributions close to that of the mainstream grains if the 
parameters for the admixture of SN material (number of SN sources,  
\N, and fraction of Si ejected by a SN, $a$) are chosen appropriately. 
In reality we have to expect a whole range of initial compositions 
reflecting different times and different degrees of homogenization 
of matter in the Galaxy. We expect that material at a given Galactic 
radius is homogenized on a time scale of less than 10$^8$ yr, the period 
of Galactic rotation. Any complete homogenization will destroy the 
local heterogeneities in which we are interested. The real local 
isotopic compositions will represent some balance between heterogeneity 
and processes of homogenization. The overall result will be the GCE of 
the elements and the isotopes, the overall trend being modified by 
local fluctuations.}

\item {3) In our model we have considered only Type II SNe of solar 
metallicity. In reality there will be a range of metallicities. 
This is for two reasons. First, we expect to encounter some range 
in age for AGB stars as will be discussed in the next section. 
Second, if local regions are highly contaminated with previous 
SN contributions, new SNe from such regions will have higher-than-average 
metallicities. According to the WW95 SNII models the addition of  
SNIIe ejecta of a given metallicity to an ISM parcel of the same 
metallicity will result in higher $^{29}$Si/$^{28}$Si and 
$^{30}$Si/$^{28}$Si ratios than those of the starting material. This 
reflects the fact that $^{29}$Si and $^{30}$Si are secondary 
isotopes. The enrichment of the heavy Si isotopes in Type II SN 
ejecta over the average ISM material has first been pointed out by 
Clayton (1988) on the basis of an idealized GCE model. The enhancement of $^{29}$Si and $^{30}$Si in the SNII 
ejecta over the starting composition is clearly seen in Fig. 11 for 
SNIIe of solar metallicity, where the average value plots to the upper right 
of the solar Si isotopic composition (this includes an assumed 
enhanced production of $^{29}$Si). However, a minimum metallicity is 
required for the contributing SNIIe in order to achieve the average Si isotopic 
ratios of the protosolar nebula or those of the mainstream grains. We 
conclude from Fig. 9 that a metallicity of \Z $>$ 0.75 $Z_{\odot}$ 
is required for the average Si isotopic composition of SNII 
ejecta to be heavier than 150 \permil, the maximum of the 
mainstream grains. This, 
however, is a lower limit since in reality low-mass 
stars do not form from pure SN ejecta.}

\item {4) It has to be clear that the Si isotopic compositions of the 
mainstream grains reflect those of their parent stars at the time of their 
birth and it is these compositions that we want to explain. However, it is 
also clear that, depending on their mass, different stars were born at 
different Galactic times, even if they all produced SiC grains at 
the same time during their AGB phase (and even this last assumption is 
not strictly valid because different SiC grains could have different IS 
life times between their formation and the birth of the solar system). 
So far we have made the implicit assumption that grains came only from 
stars in the 1.5 - 3 \msb range when we computed the inferred Si isotopic 
ratios of the mainstream grains without any AGB contributions (i.e., the 
initial isotopic ratios of the parent stars). In the following subsection
we want to explore 
this question in more detail.}

\end{itemize}

The processes leading to heterogeneity in the Si isotopic ratios 
in the ISM are much more complex than the simple mixing assumed in 
our MC model. However, and this is the most important conclusion of 
our tests, it is practically certain that the ISM at a given Galactic 
time and at a given Galactic radius is not characterized by a unique 
Si isotopic composition but by a range of compositions. This 
distribution of Si ratios will shift toward heavier isotopic 
ratios during the evolution of the Galaxy. Note that presently we do not know 
the exact distribution of the Si isotopes at a given Galactic time nor the 
relationship between Galactic time and the mean of the Si isotopic ratios 
of the distributions. For the time being we assume that the 
latter is the same 
as that of the Timmes \& Clayton (1996) model, but this point will be discussed 
in more detail in \S 4.5.

\subsection{The mass of the SiC parent stars}

In \S 3 and Figs. 5 - 8 we showed that the shift in Si isotopic 
compositions due to neutron capture in the He shell of AGB stars depends 
on stellar mass and metallicity. As we demonstrated in that section, 
the changes in the Si isotopic composition due to the $s$-process 
in AGB stars depend almost entirely on the small 
neutron exposure from the $^{22}$Ne source, with the $^{13}$C pocket 
having no influence, independent 
of the magnitude of its strength. The effects on stellar mass and 
metallicity we are discussing in this section actually are the results 
of complete stellar evolutionary calculations of the AGB phases 
using the FRANEC code. 
At solar metallicity, from a 5 \msb complete AGB evolutionary model we 
find a somewhat higher maximum temperature at the bottom of the He thermal 
pulses than in lower mass stars (1.5 \msb to 3 \ms). Because the $\alpha$-capture reaction rate is proportional to 
$T^{21}$, this higher temperature 
increases the efficiency of the neutron burst from the $^{22}$Ne 
source, correspondingly changing the predicted final Si ratios 
as illustrated in the figures.

The same tendency is found in stellar evolutionary calculations 
with the FRANEC code for AGB stars of different mass and
a metallicity of 1/3 $Z_{\odot}$, as shown in Table 4. The maximum
temperature at the bottom of the thermal pulse increases 
slightly from pulse to pulse, starting from about 
2.65 $\times 10^8$ K for the pulse 
when TDU occurs for the first time. The temperature during  the 
pulse rises in a very rapid burst; subsequently the bottom 
temperature decreases more 
or less exponentially from its maximum, with a total duration time 
(at $T >$ 2.5 $\times 10^8$ K) of a few years.

Whereas for 1.5 \msb and 3 \msb stars of 
solar metallicity the maximum shifts in $\delta$$^{30}$Si/$^{28}$Si are 
only 26 \permilb and 37 \permil, respectively, the maximum shift for a 5 \msb 
star is 180 \permil. In Figs. 14a-d we plotted the results of the MC
calculations 
for the $-$100 \permilb case if we add the shifts expected for AGB stars of 
masses 1.5 \ms, 3 \ms, and 5 \msb with solar metallicity and of 3 \msb 
with \Z = 0.006.  The ranges of shifts were added in a random, 
statistical fashion in our MC test. As can be seen, only the 1.5 \msb and 
3 \msb stars of solar metallicity give results in reasonable agreement 
with the grain data, while 5 \msb stars as well as stars with \Z = 0.006 
shift the Si isotopic compositions far to the right of the grain data 
and the solar composition. Especially for the low-metallicity case of 
Fig. 14d the predicted shifts have a much wider spread. 
It is a remarkable result of our heterogeneity model 
that, without any AGB contributions, the solar composition is one of 
the possible compositions and at the same time the mainstream data 
can be reproduced if the AGB shifts are small, as for the 1.5 \msb and 
3 \msb star models of close-to-solar metallicity. This is not true anymore if the AGB shifts are as 
large as those for 5 \msb stars.

From the above discussion, it is reasonable, even if not proven, to assume 
that mostly low-mass stars (with \M $\leq$ 3 \ms) contributed SiC to the solar system. Actually, 
there are many pieces of evidence that indicate that this is indeed the 
case:

\begin{itemize}

\item {1) Feast (1989) performed a study of the kinematics of peculiar red 
giants including S, SC, and C stars. On the basis of 427 C stars he 
estimated their mean mass to be 1.6 \ms. Although this estimate 
needs to be improved, if SiC grains came from average 
C stars, they came from low-mass stars.}

\item {2) Another argument for low masses of carbon stars is based on a 
comparison of the observed luminosities of AGB stars in the Magellanic 
clouds with predicted luminosities. Theory predicts intermediate-mass 
stars of 5 - 8 \msb to have $M_v$ of less than - 6.5 but typical  
luminosities of S and C stars are much lower, indicating low-mass stars 
(Mould \& Reid 1987;  Frogel, Mould, \& Blanco 1990; Van Loon et al. 
1998).}

\item {3) Another argument is based on theoretical predictions about the 
occurrence of hot bottom burning (HBB) in intermediate-mass (5 - 8 \ms) 
stars. HBB takes place when the bottom layers of the convective envelope 
are hot enough for some proton capture nucleosynthesis to occur. In this 
case, most \cdb dredged up from the He 
shell during the TP-AGB phase is converted to $^{14}$N, 
preventing the star from becoming a carbon 
star. There are several theoretical studies that indicate that HBB occurs in
stars of $\gtrsim$ 5 \msb of solar metallicity and in stars with 
$\gtrsim$ 4 \msb of lower metallicity (Boothroyd, Sackmann, \& Wasserburg 
1995; Forestini \& Charbonnel 1997; Lattanzio et al. 1997). For solar 
metallicity stars, the FRANEC code finds HBB in a 7 \msb but not a 5 \msb  
star. The situation is complicated by the finding that if HBB stops while 
thermal pulses and TDUs continue in a star with mass loss, the star can 
become C-rich (Frost et al. 1998; Lattanzio \& Forestini 1999). However, this happens only in 
low-metallicity stars. Furthermore, if superwinds during the advanced 
AGB phase erode the envelope quickly, it is possible that TP 
cease before the star becomes C-rich. Thus, by and large 
it is not very likely that there are a substantial number 
of C-rich intermediate-mass stars that could have contributed SiC 
to the solar system.}

\item {4) We also obtain constraints on the mass and the metallicity of the 
parent stars from the isotopic compositions measured in presolar SiC 
grains when these compositions are compared with model calculations:

\begin{itemize}

\item {i) It has already been pointed out that the heavy elements patterns 
measured in presolar SiC are well reproduced by models of neutron-capture 
nucleosynthesis in AGB stars (Gallino et al. 1997) . However, this 
agreement exists only for low-mass AGB stars of close-to-solar 
metallicity and not for intermediate-mass stars, or of AGB stars of low
metallicity. A particularly 
diagnostic isotopic ratio is the $^{96}$Zr/$^{94}$Zr ratio. The large  
$^{96}$Zr depletions measured in mainstream SiC grains (Nicolussi et al. 
1997; Pellin et al. 1999) are well reproduced only with models of AGB 
stars of 1.5 - 3 \msb and about solar metallicity (Gallino et al. 1998b), but 
higher-mass stars and stars with low metallicity are predicted to 
produce huge $^{96}$Zr excesses.} 

\item {ii) Gallino et al. (1990) pointed out that the He and Ne isotopic 
data of presolar SiC grains are best explained in terms of 
nucleosynthesis in low-mass AGB stars of close-to-solar metallicity 
(see their Fig. 1). Another important observation is that SiC grains 
do not show large $^{25}$Mg excesses (within relatively large errors). 
This again indicates low-mass stars in which $^{22}$Ne does not burn. 
Indeed, the FRANEC code yields $^{25}$Mg excesses of up to $\sim$ 200 
\permilb in the envelope of 1.5 and 3 \msb AGB stars of solar metallicity, 
whereas predicted 
excesses are an order of magnitude larger in the 5 \msb model of \Z = 
0.02 and in the 3 \msb model of \Z = 0.006.} 

\item {iii) The situation is similar with regard to the 
$^{12}$C/$^{13}$C ratios, where the observed range is best 
reproduced by low-mass AGB models of close-to-solar metallicity (Gallino et 
al. 1990;  Bazan 1991). New results from the FRANEC code confirm 
these earlier conclusions: the best agreement is obtained for 1.5 \msb 
($^{12}$C/$^{13}$C = 40 - 60) and 3 \msb ($^{12}$C/$^{13}$C = 90 - 100) AGB 
stars of solar metallicity, the 5 \msb model of solar metallicity
and the 3 \msb model of low metallicity (\Z = 
0.006) yield much higher ratios ($^{12}$C/$^{13}$C = 90 - 120 and 
100 - 700, respectively). In low-mass star models with $M \lesssim$ 2.5 \ms,
the presence of 
cold bottom processing (CBP) (Charbonnel 1995; Wasserburg et al. 1995;
Boothroyd \& Sackmann 1999) 
lowers the initial $^{12}$C/$^{13}$C ratio at the beginning of the TP-AGB 
phase.} 

\item {iv) Cold bottom processing is also important for the 
$^{14}$N/$^{15}$N ratio. The high ratios observed in many individual 
mainstream SiC grains (Fig. 1) can only be explained by CBP (Huss et al. 
1997) operating in low-mass stars.}

\end{itemize}}

\item {5) A lower limit on the masses of carbon stars can be 
obtained from models with TDU. Existing models predict TDU only for 
stars with $M \gtrsim$ 1.5 \msb (Lattanzio 1989; 
Straniero et al. 1997; Gallino 
et al. 1998a;  Busso et al. 1999b). In the FRANEC 
code the limit depends on the value of Reimer's parameter $\eta$ used. 
For a star of $M =$ 1.5 \msb of solar metallicity the limit is 
$\eta$= 0.3 for TDU to occur and for producing C/O $>$ 1 in the 
envelope during the advanced stages of the 
AGB phase. The fact that there is a 
minimum mass below which TDU does not occur is of great 
importance, since because of it SiC grains cannot originate 
from long-lived stars of low mass and low metallicity. 
It is worth noting that increasing the metallicity above 
solar works against an AGB star to become C-rich. In our 
calculations the star remains O-rich at $Z = 2 \times$ $Z_{\odot}$ 
for $M = $ 1.5 \ms, and already at $Z = 1.25 \times$ 
$Z_{\odot}$ for $M = $ 3 \ms.}

\end{itemize}

From all these considerations it appears that most presolar SiC grains 
come from AGB stars of 1.5 - 3 \msb and close-to-solar metallicity. Of 
course, it is not said that the mainstream SiC grains have to come from 
typical carbon stars. It is possible that these very large grains 
preferentially originated from stars with very dense winds and thus from
stars having masses at the upper end of the above range. However, stars 
with masses much larger than 3 \msb can definitely be excluded.

\subsection{Star lifetimes, grain lifetimes and Galactic chemical 
evolution}

Let us now return to the question of lifetimes of the possible source 
stars for SiC grains. Whereas the calculated lifetime of a 5 \msb star 
of solar metallicity is 1.1 $\times 10^8$ yr (Schaller et al. 1992), 
those of the 3 \msb and 1.5 \msb stars are 4.4 $\times 10^8$ and 
2.9 $\times 10^9$ yr, respectively. Especially the latter lifetime 
would result in a non-negligible difference in the Si isotopic ratios 
due to the overall temporal evolution of the Si isotopes in the Galaxy. 
According to the model of Timmes \& Clayton (1996) a time difference of 
2.9 $\times 10^9$ yr corresponds to a difference of 125 \permilb in the 
$^{29}$Si/$^{28}$Si and $^{30}$Si/$^{28}$Si ratios, which is almost the 
whole range covered by the mainstream grains. This means that a star that
was born 2.9 $\times 10^9$ yr before the sun should have 
$^{29}$Si/$^{28}$Si and $^{30}$Si/$^{28}$Si ratios that are, on average, 
125 \permilb smaller than the solar ratios. It also means that if stars of 
different mass and therefore different lifetimes contributed SiC grains 
to the solar system, these grains are expected to have different Si isotopic 
compositions. There are several factors 
that play a role here. One is the initial mass function for stars. There 
are more stars of lower mass and we have already pointed out that the 
estimated mean mass of carbon stars is $\sim$1.6 \msb (Feast 1989). On the other hand, 
the fact that SiC grains are relatively large 
suggests that more massive AGB stars with very dense winds, meaning 
high mass loss at low speed, were selected as the SiC grains' parent 
stars.

Let us consider two extreme cases. First we consider the case that all 
or most of the 
mainstream SiC grains came from AGB stars of approximately the same 
mass and therefore also the same time of formation. In this 
case time differences do not play a role and 
the distribution of the Si isotopic ratios of the mainstream grains 
can in principle be explained as having an origin in 
local isotopic heterogeneities 
due to the statistical nature of SN contributions to the ISM. This is 
schematically shown in Fig. 15a where the whole range of Si isotopic 
compositions exhibited by the SiC mainstream grains is interpreted 
as the spread in Si isotopes that existed at the time of formation 
of the grains' parent stars. These parent stars have to have 
approximately the same mass but at this point it is not said whether 
it is low (1.5 \ms) or high (3\ms).

The second case to be considered is one in which AGB stars of 
a {\it considerable} mass range contributed the 
mainstream grains to the solar system. In this case the 
range in Si isotopic shifts
due to the formation time difference between stars of 1.5 \msb and of 3 \msb 
is of the same order of magnitude 
as the spread of the mainstream grains. 
Variations in the Si 
isotopic ratios of individual grains are expected to arise from 
the age differences 
of their parent stars (which in turn vary because of GCE) 
and local 
heterogeneities of the Si isotopes play a complementary role. 
This situation is schematically depicted in 
Fig. 15b, where the 
spread of the mainstream grains' Si isotopic ratios is interpreted 
as a superposition of local heterogeneity distributions representative 
of different Galactic times.

Another factor that plays a role here is the lifetime of the SiC grains 
in the ISM. This lifetime has to be added to the lifetime of the AGB 
parent stars in terms of time differences between the birth of these 
stars and the formation of the solar system and the implication of 
these time differences for the Si isotopic ratios. Unfortunately, at 
present we do not have any good direct measure of grain lifetimes. 
Attempts have been made to determine IS grain lifetimes from the 
measurement of  cosmogenic $^{21}$Ne produced in the grains from the  
spallation of Si by Galactic cosmic rays (Tang \& Anders 1988b; 
Lewis et al. 1994). Estimates obtained in this way range up to 
1.3 $\times 10^8$ yr. However, besides poor knowledge of the flux of 
Galactic cosmic rays, there are many other uncertainties associated 
with this approach. Single grain measurements showed that only $\sim$ 
5\% of all SiC grains are rich in $^{22}$Ne (Nichols et al. 1991, 1992, 
1993). If one assumes that outgassing is the reason that the other 
grains lack measurable amounts of $^{22}$Ne and that the same process 
removed cosmogenic $^{21}$Ne from these grains, one arrives at much higher 
estimates for IS grain lifetimes. However, it is unclear that outgassing 
is indeed the cause for the large variations of $^{22}$Ne among single SiC 
grains. Another problem is the determination of spallation recoil 
loss from the grains. From experimental measurements of spallation 
recoil Ott \& Begemann (1997, 1999) concluded that a determination of presolar 
exposure ages from cosmogenic $^{21}$Ne is not feasible. These authors 
propose the use of spallation Xe as more promising but before this is 
done we do not have any reliable IS lifetimes for presolar grains. An 
alternative way is to use model ages derived from theoretical 
destruction rates of IS grains by SN shocks and collisions (see, e.g., 
Whittet 1992; Jones et al. 1997). Estimates range up to $\sim 10^9$ yr 
but there are also large uncertainties in this approach.

\subsection{SiC grains and the Si isotopic composition of the sun}

So far we have discussed differences in the formation time of AGB 
stars of different masses that possibly contributed SiC grains to 
the solar system. We have not discussed yet the relationship 
between the formation time of the grains' parent stars relative to 
that of the solar system and implications for their relative Si 
isotopic compositions. In the Timmes \& Clayton (1996) model the 
fact that most mainstream grains have isotopically heavier compositions 
than the solar system (implying that they are younger) 
but must have formed before the sun presents a fundamental problem. 
Our heterogeneity model alleviates this fundamental problem, because 
in principle it can explain the spread in the Si isotopic compositions 
of the mainstream grains as inhomogeneities of the Si isotopes in the 
ISM at a given time (see Fig. 15a). However, the grains' parent stars 
must have formed before the solar system and we must 
discuss the effect of this time difference on their Si isotopic 
compositions.

For the sake of discussion we again consider two extreme cases. First we consider the case that most 
mainstream SiC grains came from AGB stars of 3 \ms. The 
evolution time of such stars is 4.4 $\times 10^8$ yr. According to 
Timmes et al. (1995) and Timmes \& Clayton (1996), such a time 
difference corresponds to a shift of 
19 \permilb of the Si isotopic ratios. If we assume that the spread in 
Si isotopic ratios at the time of the birth of the parent stars 
coincides with that of the mainstream grains, the Si isotopic distribution 
at the time of solar system formation 4.4 $\times 10^8$ yr later is 
isotopically heavier by 19 \permilb (Fig. 16a). This shift is 
relatively small compared to the range of the mainstream grains. The Si 
isotopic composition of the sun, while falling at the outer edge of 
this distribution, lies still within the range of compositions 
expected to be present at the time of solar system formation. 
The fact that the sun has a 
composition that differs from those of most of the grains, is not a 
fundamental problem in this case. It simply means that the sun, 
as many other SiC 
parent stars, has an unusual composition but one that is not incompatible with expectations.

Let us next consider the other extreme, that all grains come from stars of 
1.5 \ms. The evolution time of these stars is 2.9 $\times 10^9$ yr. 
According to Timmes \& Clayton (1996) this time difference corresponds to 
a shift of the Si isotopic ratios by 125 \permilb. This means that the distributions 
of the Si isotopes at the time of star formation and at the time of solar 
system formation 2.9 $\times 10^9$ yr later are shifted by this amount 
relative to one another. This is shown in Fig. 16b where, again, we assume 
that the Si isotopic distribution of the stars coincides with that of 
the mainstream grains. This time, the inferred distribution 
2.9 $\times 10^9$ yr later is shifted so much that it seems quite
impossible 
that the solar system composition observed today can be explained as 
being part of this distribution. In other words, in the extreme case in which only stars of 
$M = $ 1.5 \msb contributed SiC grains, we are faced with the 
same fundamental problem as the Timmes \& Clayton (1996) model, namely that 
the actual solar system composition is much too light compared to the 
distribution predicted for the time of solar system formation if the mainstream grains came from old stars. 

Clayton (1997) has addressed this problem and 
has proposed a solution in terms of a systematic difference in the 
Galactic radius at which the parent stars of the mainstream grains 
on the one hand and the sun on the other hand formed. The parent stars 
are assumed to have formed at smaller Galactic radii where the 
metallicity and $^{29}$Si/$^{28}$Si and $^{30}$Si/$^{28}$Si ratios are 
believed to be higher. His model involves the diffusion of stars from 
smaller to larger Galactic radii due to scattering on IS clouds, 
following the stellar 
orbital diffusion model by Wielen, Fuchs, \& Dettbarn (1996). Such a 
model had been proposed in order to explain the spread in elemental 
abundances observed in stars. However, a more detailed quantitative 
treatment by Nittler \& Alexander (1999b) shows that with reasonable 
assumptions a diffusion model cannot account for the isotopically heavy 
Si compositions of the grains relative to the sun. Furthermore, van 
den Hoek \& de Jong (1997) pointed out that stellar orbital diffusion 
cannot sufficiently explain the elemental abundance variations. 
While we do not want to discard the orbital diffusion model, we 
hope that the heterogeneity explanation will give a more definitive answer 
to this important question. This would require a model that, 
by computing the Galactic evolution of the Si isotopes 
by taking into account incomplete mixing of 
different stellar yields, overcomes the problems discussed in 
\S 4.3 and all those connected 
to the overly simplistic nature of our approach.

It should be noted that in our estimates of the Si isotopic shifts 
associated with time differences we have used the Si isotopic evolutions 
vs. time relationship given by Timmes et al. (1995) and 
Timmes \& Clayton (1996). This relationship crucially depends on the 
relative proportion in which Type Ia and Type II SNe contribute to the 
enrichment of the ISM in Si isotopes. 
Timmes et al. (1995) attributed a dominant role to Type II 
SNe by assuming that at the time of solar system formation they 
contributed 2/3 of the Fe. Woosley et al. (1997), on the other hand 
estimated that this fraction would be 1/2. This would mean that 
the $^{28}$Si contribution from Type Ia SNe is higher and therefore 
the Si isotopes evolve more slowly toward heavier compositions. 
This in turn would mean that a Si isotopic shift corresponding to 
a given time difference (Fig. 16) is smaller than what we assumed. 

In conclusion, there are still large uncertainties as to the masses 
of the parent AGB of the grains, the ISM life times of the grains, and the 
time dependence of the evolution of the Si isotopes in the Galaxy. 
All of these uncertainties have to be clarified before we can hope to 
solve the problem of the difference of the Si isotopic compositions of 
the mainstream SiC grains and that of the solar system.

\section{Conclusions}

Mainstream SiC grains are the major group of presolar SiC grains found in 
meteorites. Although there 
is overwhelming evidence that mainstream grains have an origin in the 
expanding atmospheres of AGB stars, their Si isotopic ratios show a 
distribution (Fig. 3) that cannot be explained by nucleosynthesis in 
AGB stars. The theoretically predicted Si isotopic shifts in the 
envelope of AGB stars are either much smaller than the range observed 
in the mainstream grains (for AGB models of \M = 1.5 and 3 \msb and 
solar metallicity) or (for AGB models of \M = 5 \msb and solar 
metallicity and \M = 3 \msb and \Z  = 0.006) show a slope of $\sim$ 0.5 
correlation between the $\delta$$^{29}$Si/$^{28}$Si and 
$\delta$$^{30}$Si/$^{28}$Si values instead of the slope 1.31 correlation 
line exhibited by the grains.

The distribution of the Si isotopic ratios of the mainstream grains 
has previously been interpreted to be the result of GCE of the Si 
isotopes. In this interpretation the grains' parent stars are expected 
to have a range of different Si isotopic ratios if they were born at 
different times. In this paper we proposed an alternative explanation 
for the Si isotope distribution by invoking isotopic heterogeneities 
due to the statistical nature of the contributions of a limited number 
of SN sources to the IS material from which the grains' parent stars 
formed. The Si isotopic ratios of the ejecta of possible SN sources, 
classical Type Ia SNe, Type Ia SNe from sub-Ch white dwarfs, 
and Type II SNe of different masses, span a wide range. We developed a 
simple Monte Carlo model in which contributions from these SN sources 
were admixed in a random way to material with a given Si isotopic 
composition. As long as this composition lies on the theoretically 
expected GCE line going through the solar Si isotopic composition, 
we could show that, with the right choice of parameters, the 
distribution of the Si isotopic ratios in the mainstream grains can 
be successfully reproduced for a wide range of starting compositions. 
The parameters to be adjusted are the total number of SN sources 
selected and the fraction of the material ejected from each SN that 
is admixed to the starting material. In addition, an adjustment of the SN 
yield of $^{29}$Si by a factor of 1.5 is necessary to achieve the 
Si isotopic ratios of the solar system. 
Astronomical observations of variations of 
elemental ratios in stars are compatible with the predictions 
from our MC model.

These results demonstrate that, in principle, the mainstream distribution 
can be explained as the result of local fluctuation in the ISM due to 
the admixture of material from a limited number of SN sources to the 
preexisting IS matter. If the AGB stars that contributed SiC grains to 
the protosolar nebula were born within a short period of time (by having a 
narrow range of masses and the grains having short IS lifetimes), such 
fluctuations must have been the dominant cause of the mainstream 
distribution. If, however, the AGB parent stars had a large range of 
masses and therefore a large range of lifetimes and/or the grains themselves 
experienced a large range of residence times in the ISM, the parent 
stars must have been born at different Galactic eras and their initial 
Si isotopic ratios must show considerable variations because of the 
varying average composition of the ISM due to the GCE of the Si isotopes. 
In this case we still expect that local fluctuations will be 
superimposed on these average compositions. To simulate these complex 
processes it will be necessary to apply to the Si isotopes a
Galactic evolution model that is able to take both components properly 
into account. 
However, a successful model would require knowledge of the mass 
distribution of AGB stars that contributed SiC grains (at least in 
the size range of the single grains whose data are plotted in Fig. 3) 
and the distribution of the IS lifetimes of these grains. Both pieces 
of information are presently unknown and we can only hope that further 
progress in the study of the grains and their origin will get us 
closer to an answer.

We are grateful to Peter Hoppe for providing isotopic data on the 
Murchison KJE size fraction, to Gary Huss for providing his Orgueil 
data, and to Stan Woosley for providing unpublished results of the Si 
yields from his and Weaver's \Z = 0.5 $Z_{\odot}$ and 
\Z = 2 $Z_{\odot}$ supernova models. 
We are deeply indebted to Oscar Straniero, Maurizio Busso, Alessandro 
Chieffi and Marco Limongi for all 
their scientific input and thank Don Clayton for ideas and discussions. 
The detailed and thoughtful review by Don Clayton substantially 
contributed to the final version of this paper. 
ML gratefully acknowledges the invaluable help of John Lattanzio. 
EZ deeply appreciates the hospitality extended to him by Roberto Gallino
during a visit to the Dipartimento di Fisica Generale of the University 
of Torino and the support for this visit provided by the Gruppo Nazionale di
Astronomia del CNR. SA acknowledges the support for a visit to the same 
Department provided by the University of Torino.
This work was supported by an Overseas Postgraduate Research Scheme 
award (ML), NASA grant NAG5-8336 (SA and EZ) and by MURST Cofin98 
Progetto Evoluzione Stellare (RG).

\newpage
\appendix{\bf Appendix: Monte Carlo calculations}

The problem at hand is to calculate the Si isotopic composition of material 
that is obtained by mixing contributions from a {\it limited} number of SN
sources of different types to material with a given isotopic composition 
and to determine whether the isotopic compositions observed in mainstream 
SiC grains can be reproduced in this way. If the SN contributions, whose 
individual Si isotopic compositions vary from one SN to the next, are 
admixed to the starting material in a random way, each of the resulting 
final mixtures will have a different isotopic composition because of the 
stochastic nature of this process; we will obtain a distribution of 
compositions. It is quite natural to apply Monte Carlo (MC) methods to 
generate a sample of such mixtures whose isotopic distribution can be 
compared with the isotopic distribution of the SiC grains.

In the MC calculations we considered as starting material one solar mass 
with solar elemental abundances (Anders \& Grevesse 1989). One \msb is a 
convenient mass unit since it is of the same order of magnitude as the
masses of the AGB stars that contributed SiC grains to the solar system. 
We performed several series of MC calculations with different initial Si
isotopic ratios in each case. For each such series we started with a 
fixed initial Si isotopic composition. To this starting material we added 
contributions from \Nb different SN sources that were selected at random
from a list of candidates. As SN sources we considered Type II SNe of
solar metallicity, Type Ia SNe of the W7 model and the sub-Chandrasekhar 
model. We assumed that on average Type II SNe represented 80\%, Type Ia
SNe of the W7 model 12\% and Type Ia SNe of the sub-Chandrasekhar model 8\%
of the total number of SN sources contributing Si to the mix. These 
relative frequencies are those recommended by Cappellaro et al. (1997).
This means that if we added contributions from \Nb = 100 SNe, {\it on
average}, 80 of these would be of Type II, 12 of Type Ia and 8 
sub-Chandrasekhar SNe. However, because of the statistical nature of 
this mix the exact numbers would vary from one mixture to the next.

Among the Type II SN models we selected SNe of different masses which 
were weighted according to a Salpeter initial mass function
$f(M)dM = M^{-2.35} dM$. We chose a mass range of 10.5 to 42.5 \ms, which
is the mass range covered by the Type II SN models of solar metallicity 
of WW95 (Table 3). Within this mass range, a mass was chosen according 
to the probability of the initial mass function and the SN model of WW95 
was selected whose mass was closest to the chosen mass. Thus for masses 
between 10.5 and 11.5 we took the $M$  = 11 \msb model from WW95, for
masses between 11.5 and 12.5 the $M$ = 12 \msb model and so on. The final
mixture thus contained contributions from both kinds of Type Ia SNe and 
from Type II SNe of different masses but, as already mentioned, each mix 
contained a different combination of these sources.

For each SN, the total ejected mass of each of the three Si isotopes was 
multiplied by a constant factor $a$ and then added to the starting 
composition. The final amount of a given isotope $i$ is given as the 
mass fraction $X^i = X^{i}_{initial} + \Sigma_{1}^{N} a \times 
M^{i}_{ejected}(SN)$. Thus $N \times a$ is a measure of the total amount
of material added to the starting material. Since we would like the Si 
isotopic composition of the mix to match the compositions of the SiC 
mainstream grains, we first computed the mean value and standard 
deviation of the $\delta$$^{29}$Si/$^{28}$Si and
$\delta$$^{30}$Si/$^{28}$Si values of the mainstream grains. If we 
restrict ourselves to the high-quality data plotted in Fig. 3 we obtain:

$\delta$$^{29}$Si/$^{28}$Si$_{mean}$ = 50.4 \permil; 
$\delta$$^{29}$Si/$^{28}$Si$_{std.dev}$ = 42.7 \permil

$\delta$$^{30}$Si/$^{28}$Si$_{mean}$ = 52.0 \permil
$\delta$$^{30}$Si/$^{28}$Si$_{std.dev}$ = 30.6 \permil

Since the grain data contain also the nucleosynthetic contributions of 
the AGB parent stars, we subtracted $\delta$$^{29}$Si/$^{28}$Si$_{AGB}$ =
20 \permilb and $\delta$$^{29}$Si/$^{28}$Si$_{AGB}$ = 25 \permil, the
average of the shifts predicted for 1.5 and 3 \msb AGB stars (Fig. 5).
This gives $\delta$$^{29}$Si/$^{28}$Si$_{mean}$ = 30.4 \permilb and 
$\delta$$^{30}$Si/$^{28}$Si$_{mean}$ = 27 \permilb for the mean
compositions of the parent stars before their Si isotopes were affected by
neutron capture during the AGB phase.

The next question is how these grain averages can be obtained by our 
mixing procedure. The isotopic composition of the mix is expected to 
depend on the relative proportion of the amount of starting 
material to the amount of SN contributions. Note that in this discussion 
we understand under mix the exact mix between the starting composition 
and the weighted average of the SN contributions. This mix has a unique 
composition and does not suffer from the statistical fluctuations caused 
by a stochastic admixture of individual SN contributions. Alternatively, 
it can be considered to be the average over a large number of statistical 
mixtures. If the SN contributions are small, the final Si isotopic ratios
will be very similar to those of the starting material. If the SN 
contributions dominate, the final ratios will approach those of the 
weighted average of all different SN sources. On a Si 3-isotope plot 
the Si isotopic composition will move along a straight line from that 
of the starting material to that of the SN average as the ratio of SN to 
starting material is increased from zero to infinity (Fig. 17). Note that 
in Fig. 17 the SiC mainstream grain average has been corrected for the 
\s-process contributions of the AGB stars.

For some value of this mixing ratio (or, in other words, for some value of  
$N \times a$) the isotopic composition of the mix will match the average
of the grain data. However, this can only be the case if in a 3-isotope 
plot the starting composition, the average grain composition, and the 
average SN composition lie on a straight line. This condition is not 
satisfied if we choose as a starting composition one in agreement with 
the expected Galactic evolution of the Si isotopes, i.e. a composition 
on a straight line connecting the origin with the solar isotopic 
composition (see Fig. 11). The reason is that the average isotopic 
composition of all SN sources lies far below this line (see Fig. 17). 
This has been discussed in detail in \S 4.1 and, following Timmes 
\& Clayton (1996), we argued for an adjustment of the $^{29}$Si
production rate by a factor \fb = 1.5 so that the main Galactic evolution 
of the Si isotopes goes through the solar composition. Although the 
average Si isotopic ratios of the mainstream grains corrected for 
AGB contributions ($\delta$$^{29}$Si/$^{28}$Si$_{mean}$ = 30.4 \permilb 
and $\delta$$^{30}$Si/$^{28}$Si$_{mean}$ = 27 \permil) slightly deviate 
from the Galactic evolution line through the solar composition (as defined 
by $\delta$$^{29}$Si/$^{28}$Si = $\delta$$^{30}$Si/$^{28}$Si), this
difference is negligible and the factor \fb = 1.5 yields a satisfactory
alignment of starting composition, grain composition and SN average for 
all selected starting compositions (one of them is shown in Fig. 17).

\newpage
With this adjustment factor we can next determine the value of $N \times 
a$ (i.e. the mixing ratio between starting material and SN contributions) 
for which the mix matches the average of the SiC grains. For the example 
in Fig. 17 we obtain $N \times a$ = 0.0049. We will obtain an exact value
for $N \times a$ only if we use the weighted average of the SN
contributions or in the limit of very large values for \N. For a limited
number of SN sources, the Si isotopic composition of the mix will vary 
from one mixture to the next and the spread in the isotopic ratios is 
expected to vary with $1/\sqrt{N}$. We thus can determine the number of SN
sources whose admixture will result in a spread of Si isotopic ratios
that is commensurate with the spread of the grain data. For this purpose 
we generated a total number of 200 mixtures. This is an arbitrary number
but is comparable to the number of data points for the SiC mainstream
grains (Fig. 3). For these 200 cases we calculated the standard deviation 
from the mean of the Si isotopic ratios and determined the number \Nb
of SN sources for which these standard deviations match those of the SiC
mainstream data. Strictly speaking, we would obtain an exact solution for the 
best fit only for an infinite number of cases. Because of the statistical
fluctuations associated with a finite number of cases, one set of 200
cases will yield a different best-fit value for \Nb than another set. In
other words, for a limited number of cases we obtain a certain range for
the parameter \Nb (and consequently of $a$) that yields a good fit. In
summary, the requirement that the distribution of Si isotopic compositions
generated by the MC mixing model matches that of the SiC mainstream data
allows us to determine the parameters \Nb and $a$, the number of SN
sources that contribute to the final mixture and the fraction of material 
from each. As can be clearly seen from Fig. 17, the average Si isotopic
composition of SNe of solar metallicity, adjusted for an increased 
$^{29}$Si yield, is heavier than that of the solar system and of the SiC
mainstream grains and the addition of SN material will, in general,
shift the composition of the mixture toward larger $^{29}$Si/$^{28}$Si 
and $^{30}$Si/$^{28}$Si ratios. As can be seen from Fig. 10 and Table 3,
it is mostly the contributions of the Type II SNe of 30, 35 and 40 \msb 
that contribute to this isotopically heavy composition. However, because 
of the form of the initial mass function, the frequencies of these 
high-mass SNe are low. The probability for a 35 \msb SN to be among the 
mix considered for our MC calculation is only 3.6\% and that for a 40 \msb
SN is 2.6\%. If the total number of sources \Nb is not large, as in the
case shown in Fig. 12b, for which \Nb = 70, it is possible that in a few 
cases the SN contributions can actually be isotopically light so that the
resulting mixture plots to the lower left of the starting composition 
(which is solar in the case of  Fig. 12b). This is not the case anymore
if \Nb is larger such as for the examples shown in Figs. 12c and 12d,
where all the mixtures are isotopically heavier than the starting 
compositions, $\delta$$^{29}$Si/$^{28}$Si = $\delta$$^{30}$Si/$^{28}$Si 
= $-$100 and $-$200 \permil, respectively.

\newpage
\centerline{\bf FIGURE CAPTIONS}

\setcounter{figure}{0}

\figcaption[papsi_1]{Nitrogen and carbon isotopic ratios measured
by secondary ion mass spectrometry in individual presolar SiC grains. 
Five different classes can be distinguished on the basis of the C, N, and 
Si isotopic ratios (see also Fig. 2). The abundances of the different 
classes among all meteoritic SiC are indicated. Note that the numbers of 
grains of different types in the plot do not correspond to their observed 
frequencies but that grains from rare classes have been selectively 
located by ion imaging and are thus over-represented in the graph. Data are
from Hoppe et al. (1994, 1996a,b), Nittler et al. (1995), Gao et al. (1996), 
Amari et al. (1997a,b), and Gao \& Nittler (1997).\label{fig1}} 

\figcaption[papsi_2]{Silicon isotopic ratios measured in individual presolar 
SiC grains. Isotopic ratios are plotted as $\delta$-values, permil (\permil)
deviations from the solar system ratios: 
$\delta$$^{i}$Si/$^{28}$Si = [($^{i}$Si/$^{28}$Si)$_{meas}$/
($^{i}$Si/$^{28}$Si)$_{\odot} -$ 1] $\times$ 1000. 
In this notation the solar ratios have $\delta$-values of zero. The same 
classes as those in Fig. 1 are indicated in the figure. Data are from 
the same sources as in Fig. 1.\label{fig2}} 

\figcaption[papsi_3]{Silicon isotopic ratios of ``mainstream'' SiC 
grains plotted as $\delta$-values 
from the meteorites Murchison (CM2) and Orgueil (CI). The grain 
data plot along a line with slope 1.31, which is indicated in the graph. 
Data are from Hoppe et al. (1994, 1996a) and Huss et al. (1997).
\label{fig3}}
 
\figcaption[papsi_4]{The Ti isotopic ratios measured in single 
mainstream SiC grains from the meteorite Murchison are plotted against 
their $^{29}$Si/$^{28}$Si ratios. The Ti ratios are expressed as 
$\delta$-values as are the Si ratios (see Fig. 2). There is 
a reasonably good correlation between the Ti ratios, especially the  
$^{46}$Ti/$^{48}$Ti and $^{47}$Ti/$^{48}$Ti ratios, and the Si 
ratios. This indicates that both the Si and Ti isotopic compositions 
are affected by the same process. Circles are data from Hoppe et al. 
(1994), squares from Alexander \& Nittler (1999).\label{fig4}} 

\figcaption[papsi_5]{Silicon and Ti isotopic ratios (plotted as 
$\delta$-values) predicted for 
the envelope of AGB stars of 1.5 and 3 \msb of solar metallicity during 
dredge-up of material from the He intershell in the TP phase. Solar 
isotopic ratios were assumed as the starting composition. Small solid 
symbols indicate that the stars' envelopes have O $>$ C, large symbols 
are used when the stars turn into carbon stars (C $>$ O). Plotted are 
the results for three different choices of the \ctb pocket, ST 
indicating the standard case of Gallino et al. (1998), while in the 
cases d3 and u2 the \ctb mass fraction is one third and twice the 
amount of the standard case. As can be seen, the Si isotopic ratios 
are essentially independent of the amount of \ct, whereas this is not 
the case for Ti. This means that neutron-capture nucleosynthesis of Si 
is mostly affected by the $^{22}$Ne and not by the \ctb neutron 
source. In the Si isotope plot on the upper left the correlation 
line of the mainstream grains is indicated. 
A comparison of the predicted Si isotopic ratios with the ratios 
measured in mainstream grains (Fig. 3) shows that the grain Si data 
cannot be explained by nucleosynthesis in AGB stars.\label{fig5}} 

\figcaption[papsi_6]{Titanium isotopic ratios predicted for AGB envelopes of 
1.5 and 3 \msb stars of solar metallicity 
are plotted against the $^{29}$Si/$^{28}$Si ratio. 
All ratios are plotted as $\delta$-values.
Symbols are the same as those in Fig. 5. A comparison of these graphs 
with the grain data plotted in Fig. 4 shows that the spread in the Si 
and Ti isotopic compositions of the grains cannot be explained by 
nucleosynthesis in AGB stars. \label{fig6}} 

\figcaption[papsi_7]{Silicon and Ti isotopic ratios predicted for the 
envelope of a 5 \msb star of solar metallicity. All ratios are plotted as
$\delta$-values. Symbols are the same as those in Fig. 5. While 
the maximum $\delta$$^{29}$Si/$^{28}$Si values reach those observed in 
mainstream grains (
Fig. 3), $\delta$$^{30}$Si/$^{28}$Si values are 
already $\sim$ 120 \permilb when the star becomes C-rich. Furthermore, the slope 
of the correlation line of the theoretical ratios is very different 
from that of the mainstream line plotted in the left upper graph. 
\label{fig7}} 

\figcaption[papsi_8]{Silicon and Ti isotopic ratios (plotted as
$\delta$-values) predicted for the 
envelope of a 3 \msb star with lower than-solar metallicity (\Z  = 0.006).
Symbols are the same as those in Fig. 5. The Si 3-isotope correlation 
line of the theoretical ratios has a slope of $\sim$ 0.5, much smaller than 
that of the mainstream data (upper left graph). \label{fig8}} 

\figcaption[papsi_9]{Silicon isotopic ratios plotted as $\delta$-values
of the average ejecta of Type II 
SN models as function of metallicity. The yields of SNIIe of different 
masses were averaged according to a frequency distribution of $M^{-2.35}$ 
per unit mass interval. Silicon yields of Type II SNe of metallicity 0.1 
$Z_{\odot}$ and $Z_{\odot}$ are from WW95, those for 0.5 $Z_{\odot}$ and 2 
$Z_{\odot}$ from unpublished data by Weaver \& Woosley.\label{fig9}} 

\figcaption[papsi_10]{Average Si isotopic ratios of the ejecta of Type 
II SN models of different metallicity (see Fig. 9) are plotted as
$\delta$-values together 
with predictions for individual Type II SNe of different masses for the 
0.1 $Z_{\odot}$ and the $Z_{\odot}$ cases (WW95). Also plotted are 
the predictions of the W7 Type I SN model (Thielemann et al. 1986, Nomoto 
et al. 1997) and the Type Ia SN model based on a sub-Ch white dwarf (Woosley 
\& Weaver 1994). For a given metallicity there is a large spread 
in the Si isotopic ratios predicted for stars of different masses. For 
\Z = $Z_{\odot}$ most stars with \M $<$ 25 \msb have ratios smaller 
than the solar ratios whereas stars with  \M $\geq$ 30 \msb have much 
larger ratios.\label{fig10}} 

\figcaption[papsi_11]{Silicon isotopic ratios of mainstream grains plotted
as $\delta$-values and 
compared with those predicted for the ejecta from the WW95 Type II 
SN models of solar metallicity. For these models we plotted the predicted 
values for individual mass stars with \M $\leq$ 25 \msb (the stars with 
\M $\geq$ 30 \msb plot outside of the boundary of the graph) as well 
as the averages of SNe with \M $\leq$ 25 \ms, those with \M $\geq$ 30 
\ms, and of all Type II SNe. The $^{29}$Si isotopic yields of all Type II SN 
models were adjusted (multiplied by a factor of 1.5) so that the 
$^{29}$Si/$^{30}$Si ratio of the average of all SNe is solar 
(i.e., in the graph the point for the average lies on a straight line 
through the solar isotopic composition).\label{fig11}} 

\figcaption[papsi_12]{Silicon isotopic ratios measured in 
mainstream SiC grains and plotted as $\delta$-values (open squares) 
are compared with the results of 
Monte Carlo calculations (filled circles) in which contributions from 
different SN sources are added to a starting isotopic composition in a 
random way (see text for details). In each case 200 compositions are 
plotted. (a) Starting composition: average of the mainstream grain 
ratios. Starting amount: 1 \msb of material with solar elemental 
abundances. Contributions from \Nb = 100 SNe were randomly added and 
subtracted from this starting material. The fraction taken from each 
SN source was $a$ = 1.5 $\times 10^{-5}$. 
(b) Starting composition is solar; \Nb  = 70; $a$ = 1.7 $\times 10^{-5}$. 
(c) Starting composition:  $\delta$$^{29}$Si/$^{28}$Si = 
$\delta$$^{30}$Si/$^{28}$Si = $-$ 100 \permil;  \Nb  = 420; $a$ =
1.1 $\times 
10^{-5}$. (d) Starting composition: $\delta$$^{29}$Si/$^{28}$Si = 
$\delta$$^{30}$Si/$^{28}$Si = $-$ 200 \permil; \Nb = 600; $a$ = 1.5 
$\times 
10^{-5}$. \label{fig12}} 

\figcaption[papsi_13]{(a) Elemental ratios calculated from the Monte Carlo 
model whose Si isotopic ratios are shown in Fig. 12b. The random nature 
of the SN contributions results not only in a range of Si isotopic 
compositions but also in a range of elemental ratios. The correspondent  
[Fe/H] varies from $\sim$ 0.06 to $\sim$ 0.1. (b) Elemental ratios
observed in stars (Edvardsson et al. 1993). We selected only stars
with -0.1 $<$ [Fe/H] $<$ 0.1.\label{fig13}} 
  
\figcaption[papsi_14]{Silicon isotopic ratios measured in mainstream SiC 
grains and plotted as $\delta$-values 
(open squares) are compared with the Monte Carlo results shown in 
Fig. 12c (starting composition $\delta$$^{29}$Si/$^{28}$Si = 
$\delta$$^{30}$Si/$^{28}$Si = $-$ 100 \permil) if the isotopic shift
predicted for different AGB models (Figs. 5, 7, and 8) are
added. The whole range of shifts predicted under the condition 
C $>$ O was added in a random way. (a) AGB star of 1.5 \msb with solar 
metallicity. (b) AGB star of 3 \msb with solar metallicity. (c) AGB star of 
5 \msb with solar metallicity. (d) AGB star of 3 \msb with \Z = 0.006.
\label{fig14}} 

\figcaption[papsi_15]{Schematic representation of the Si isotopic 
ratios of the SiC mainstream grains. (a) If it is assumed that the 
grains' parent stars have the same mass and were therefore born at 
approximately the same time, the spread of the mainstream grain 
compositions can be interpreted as being due to the heterogeneity 
of the Si isotopes in the ISM at the formation time of grains' 
parent stars. (b) If the grains come from AGB stars with a range 
of masses, then the spread of the mainstream grain 
compositions can be interpreted as a superposition of Si isotopic 
compositions of the ISM at different Galactic ages. In this case, 
both, isotopic heterogeneity and the Galactic evolution of 
the Si isotopes contribute to the distribution observed in the 
grains.\label{fig15}} 

\figcaption[papsi_16]{Schematic representation of the relative 
Si isotopic distributions at two different Galactic times. (a) In this 
case the time difference is 4.4 $\times 10^8$ yr, the evolution 
time of stars of 3 \msb mass. The distribution at 
time $T_{1}$ is assumed to be that of the SiC mainstream grains. 
4.4 $\times 10^8$ yr later, at the time of solar system formation, 
the distribution is shifted by 19 \permilb. This shift is so small that 
the actually measured composition of the sun is compatible with the 
expected spread. (b) For 
a time difference of 2.9 $\times 10^9$ yr, the evolution 
time of stars of 1.5 \msb mass, the predicted isotopic shift 
(125 \permilb) is so large, that the sun's composition is incompatible 
with the expected spread.\label{fig16}} 

\figcaption[papsi_17]{In a Si 3-isotope plot the Si isotopic 
composition of a mixture of two components lies on a straight line
connecting these two components. From this it follows
that, if we want to reproduce the average isotopic composition 
of the mainstream SiC grains as a mixture of an arbitrary starting 
composition (here assumed to be $\delta$$^{29}$Si/$^{28}$Si =
$\delta$$^{30}$Si/$^{28}$Si = $-$100 \permil) and the average of all 
SN contributions, the average SN composition has to lie on a straight line
connecting the starting and the average grain compositions. However, 
the weighted average of all SN sources lies far below this line. 
An adjustment of the $^{29}$Si yield from SN sources by a factor
\fb = 1.5 is necessary to achieve agreement. Note that the SiC grain
average has been corrected for the \s-process contributions predicted for
the AGB parent stars of the grains. \label{fig17}} 

\end{document}